\documentclass[12pt]{article}
\usepackage{amssymb,amscd,array}
\catcode `\@=11
\@addtoreset{equation}{section}

\newtheorem{thm}{Theorem}[section]

\newcommand{\be}{\begin{equation}}
\newcommand{\ee}{\end{equation}}
\newcommand{\bea}{\begin{eqnarray}}
\newcommand{\eea}{\end{eqnarray}}

\newcommand{\N}{\mathbb{N}}
\newcommand{\C}{\mathbb{C}}

\begin{document}
\begin{titlepage}

\begin{center}
{\bf \Large{Super-Renormalizablity of Yang-Mills Models in the Third Order of Perturbation
Theory\\}}
\end{center}
\vskip 1.0truecm
\centerline{D. R. Grigore, 
\footnote{e-mail: grigore@theory.nipne.ro}}
\vskip5mm
\centerline{Department of Theoretical Physics}
\centerline{ Institute for Physics and Nuclear Engineering ``Horia Hulubei"}
%\centerline{Institute of Atomic Physics}
\centerline{Bucharest-M\u agurele, P. O. Box MG 6, ROM\^ANIA}

\vskip 2cm
\bigskip \nopagebreak
\begin{abstract}
\noindent
We continue the investigation from a previous paper concerning the super-renormalizablity
of gauge models going to the third order of the perturbation theory. Here we consider 
only the Yang-Mills case and we prove that this property is true iff some supplementary
restrictions are imposed on the constants appearing in the interaction Lagrangian.
The usual standard model does not verify these restrictions, but there is hope that
such models do exist and they are in agreement with the phenomenology. We
consider here only the even-parity contributions.
\end{abstract}
%\newpage\setcounter{page}1
\end{titlepage}

\section{Introduction}

The general framework of perturbation theory consists in the construction of
some
distribution-valued operators called chronological products \cite{BS}. We prefer
the 
framework from \cite{DF}: for every set of Wick monomials 
$ 
W_{1}(x_{1}),\dots,W_{n}(x_{n}) 
$
acting in some Fock space
$
{\cal H}
$
one associates the distribution-valued operator
$ 
T(W_{1}(x_{1}),\dots,W_{n}(x_{n})) \equiv
T^{W_{1},\dots,W_{n}}(x_{1},\dots,x_{n})
$
such that a set of axioms, essentially proposed by Bogoliubov, are verified.
The modern construction of the chronological products can be done recursively
according to Epstein-Glaser prescription \cite{EG}, \cite{Gl} (which reduces the induction
procedure to a distribution splitting of some distributions with causal support)
or according to Stora prescription \cite{PS} (which reduces the renormalization
procedure to the process of extension of distributions). These products are not
uniquely defined but there are some natural limitation on the arbitrariness. If
the arbitrariness does not grow with $n$ we have a renormalizable theory. An
equivalent point of view uses retarded products \cite{St1}.

The description of higher spins in the perturbation theory can be problematic.
If we describe them by fields carrying only physical degrees of freedom, then 
the theories are usually not renormalizable. However, one can save renormalizability 
using ghost fields. Such theories are defined in a Fock space
$
{\cal H}
$
with indefinite metric, generated by physical and un-physical fields (called
{\it ghost fields}). One selects the physical states assuming the existence of
an operator $Q$ called {\it gauge charge} which verifies
$
Q^{2} = 0
$
and such that the {\it physical Hilbert space} is by definition
$
{\cal H}_{\rm phys} \equiv Ker(Q)/Im(Q).
$
The fact that two distinct mathematical states from 
$
{\cal H}
$
can be associated to the same physical context is called {\it gauge freedom} and the 
corresponding theories are called {\it gauge theories}.
The graded commutator
$
d_{Q}
$
of the gauge charge with any operator $A$ of fixed ghost number
\be
d_{Q}A = [Q,A]
\ee
(where
$
[\cdot,\cdot]
$
denotes the graded commutator) verifies
\be
d_{Q}^{2} = 0
\ee
so
$
d_{Q}
$
is a co-chain operator in the space of Wick polynomials. 
 
A gauge theory assumes also that there exists a Wick polynomial of null ghost
number
$
T(x)
$
called {\it the interaction Lagrangian} such that
\be
d_{Q}T = i \partial_{\mu}T^{\mu}
\label{gau1}
\ee
for some other Wick polynomials
$
T^{\mu}.
$
This relation means that the expression $T$ leaves invariant the physical
states, at least in the adiabatic limit. Indeed, we have:
\be
T(f)~{\cal H}_{\rm phys}~\subset~~{\cal H}_{\rm phys}  
\label{gau2}
\ee
up to terms which can be made as small as desired (making the test function $f$
flatter and flatter). In all known models one finds out that there exist a chain
of Wick polynomials
$
T^{\mu},~T^{[\mu\nu]},~T^{[\mu\nu\rho]},\dots
$
such that:
\be
d_{Q}T = i \partial_{\mu}T^{\mu}, \quad
d_{Q}T^{\mu} = i \partial_{\nu}T^{[\mu\nu]}, \quad
d_{Q}T^{[\mu\nu]} = i \partial_{\rho}T^{[\mu\nu\rho]},\dots
\label{descent}
\ee
where the brackets emphasize completely antisymmetric in all indexes; it follows that 
the chain of relation stops after a finite number of steps. We can also use
a compact notation
$
T^{I}
$
where $I$ is a collection of indexes
$
I = [\nu_{1},\dots,\nu_{p}]~(p = 0,1,\dots,)
$
and one can write compactly the relations (\ref{descent}) as follows:
\be
d_{Q}T^{I} = i~\partial_{\mu}T^{I\mu}.
\label{descent1}
\ee
All these polynomials have the same canonical dimension
\be
\omega(T^{I}) = \omega_{0},~\forall I
\ee
and the ghost number:
\be
gh(T^{I}) = |I|.
\ee
If the interaction Lagrangian $T$ is Lorentz invariant, then one can prove that
the expressions
$
T^{I},~|I| > 0
$
can be taken Lorentz covariant.

Now we can construct the chronological products
\be
T^{I_{1},\dots,I_{n}}(x_{1},\dots,x_{n}) \equiv
T(T^{I_{1}}(x_{1}),\dots,T^{I_{n}}(x_{n}))
\ee
according to the recursive procedure. We say that the theory is gauge invariant
in all orders of the perturbation theory if the following set of identities
generalizing (\ref{descent1}):
\be
d_{Q}T^{I_{1},\dots,I_{n}} = 
i \sum_{l=1}^{n} (-1)^{s_{l}} {\partial\over \partial x^{\mu}_{l}}
T^{I_{1},\dots,I_{l}\mu,\dots,I_{n}}
\label{gauge}
\ee
are true for all 
$n \in \N$
and all
$
I_{1}, \dots, I_{n}.
$
Here we have defined
\be
s_{l} \equiv \sum_{j=1}^{l-1} |I|_{j}.
\ee

Such identities can be usually broken by {\it anomalies} i.e. expressions of the
type
$
A^{I_{1},\dots,I_{n}}
$
which are quasi-local and might appear in the right-hand side of the relation
(\ref{gauge}). It still an unsolved problem to prove, at least in the causal
formalism, that the anomalies can be eliminated by convenient redefinitions
of the chronological products.

If one can choose the chronological products such that gauge invariance is true
then  there is still some freedom left for redefining them. To be able to decide
if the theory is renormalizable one needs the general form of such
arbitrariness. 

In a recent paper \cite{super2} we have proved that we have some super-renormalizablity
properties for loop contributions, in the second order of the perturbation theory. 
We remind the relevant cohomology terminology. We consider a {\it cochains} to be
an ensemble of distribution-valued operators of the form
$
C^{I_{1},\dots,I_{n}}(x_{1},\dots,x_{n}),~n = 1,2,\cdots
$
(usually we impose some supplementary symmetry properties) and define the
derivative operator $\delta$ according to
\be
(\delta C)^{I_{1},\dots,I_{n}}
= \sum_{l=1}^{n} (-1)^{s_{l}} {\partial\over \partial x^{\mu}_{l}}
C^{I_{1},\dots,I_{l}\mu,\dots,I_{n}}.
\ee
We can prove that 
\be
\delta^{2} = 0.
\ee
Next we define
\be
s = d_{Q} - i \delta,\qquad \bar{s} = d_{Q} + i \delta
\ee
and note that
\be
s \bar{s} = \bar{s} s = 0.
\ee
We call {\it relative cocycles} the expressions $C$ verifying
\be
sC = 0
\ee
and a {\it relative coboundary} an expression $C$ of the form
\be
C = \bar{s}B.
\ee
The relation (\ref{gauge}) is simply the cocycle condition
\be
sT = 0
\ee
and we have showed that the loop contributions of the second order of the perturbation
theory are coboundaries, up to super-renormalizable contributions.

In this paper we consider only Yang-Mills models and extend the result to the third
order of the perturbation theory. We will prove that this can be done if we impose
some supplementary restrictions on the various constants appearing in the interaction
Lagrangian. It seems that the usual standard model of electro-weak and strong interactions
does not verify these supplementary restrictions, but we hope that one can find an
alternative model, verifying these restrictions and compatible with the phenomenology.

In the next Section we will briefly present the Yang-Mills model in our preferred compact
notations. In Section \ref{causal} we give the basic ideas of causal
perturbation theory. In Section \ref{sr} we present our main result concerning
super-renormalizability in the third order of the perturbation theory for the
Yang-Mills model.
\newpage
\section{Yang-Mills Models\label{ym}}
We give some results from \cite{cohomology}.
\subsection{Massless Particles of Spin $1$ (Photons)}

We consider a vector space 
$
{\cal H}
$
of Fock type generated (in the sense of Borchers theorem) by the vector field 
$
v_{\mu}
$ 
(with Bose statistics) and the scalar fields 
$
u, \tilde{u}
$
(with Fermi statistics). The Fermi fields are usually called {\it ghost fields}.
We suppose that all these (quantum) fields are of null mass. Let $\Omega$ be the
vacuum state in
$
{\cal H}.
$
In this vector space we can define a sesquilinear form 
$<\cdot,\cdot>$
in the following way: the (non-zero) $2$-point functions are by definition:
\bea
<\Omega, v_{\mu}(x_{1}) v_{\mu}(x_{2})\Omega> =i~\eta_{\mu\nu}~D_{0}^{(+)}(x_{1}
- x_{2}),
\nonumber \\
<\Omega, u(x_{1}) \tilde{u}(x_{2})\Omega> =- i~D_{0}^{(+)}(x_{1} - x_{2})
\qquad
<\Omega, \tilde{u}(x_{1}) u(x_{2})\Omega> = i~D_{0}^{(+)}(x_{1} - x_{2})
\eea
and the $n$-point functions are generated according to Wick theorem. Here
$
\eta_{\mu\nu}
$
is the Minkowski metrics (with diagonal $1, -1, -1, -1$) and 
$
D_{0}^{(+)}
$
is the positive frequency part of the Pauli-Jordan distribution
$
D_{0}
$
of null mass. To extend the sesquilinear form to
$
{\cal H}
$
we define the conjugation by
\be
v_{\mu}^{\dagger} = v_{\mu}, \qquad 
u^{\dagger} = u, \qquad
\tilde{u}^{\dagger} = - \tilde{u}.
\ee

Now we can define in 
$
{\cal H}
$
the operator $Q$ according to the following formulas:
\bea
~[Q, v_{\mu}] = i~\partial_{\mu}u,\qquad
[Q, u] = 0,\qquad
[Q, \tilde{u}] = - i~\partial_{\mu}v^{\mu}
\nonumber \\
Q\Omega = 0
\label{Q-0}
\eea
where by 
$
[\cdot,\cdot]
$
we mean the graded commutator. One can prove that $Q$ is well defined: basically it leaves
invariant the causal commutation relations.  The usefulness of this construction follows 
from:
\begin{thm}
The operator $Q$ verifies
$
Q^{2} = 0.
$ 
The factor space
$
Ker(Q)/Ran(Q)
$
is isomorphic to the Fock space of particles of zero mass and helicity $1$
(photons). 
\end{thm}

%\newpage

\subsection{Massive Particles of Spin $1$ (Heavy Bosons)}

We repeat the whole argument for the case of massive photons i.e. particles of
spin $1$ and positive mass. 

We consider a vector space 
$
{\cal H}
$
of Fock type generated by the vector field 
$
v_{\mu},
$ 
the scalar field 
$
\Phi
$
(with Bose statistics) and the scalar fields 
$
u, \tilde{u}
$
(with Fermi statistics). We suppose that all these (quantum) fields are of mass
$
m > 0.
$
In this vector space we can define a sesquilinear form 
$<\cdot,\cdot>$
in the following way: the (non-zero) $2$-point functions are by definition:
\bea
<\Omega, v_{\mu}(x_{1}) v_{\mu}(x_{2})\Omega> =i~\eta_{\mu\nu}~D_{m}^{(+)}(x_{1}
- x_{2}),
\quad
<\Omega, \Phi(x_{1}) \Phi(x_{2})\Omega> =- i~D_{m}^{(+)}(x_{1} - x_{2})
\nonumber \\
<\Omega, u(x_{1}) \tilde{u}(x_{2})\Omega> =- i~D_{m}^{(+)}(x_{1} - x_{2}),
\qquad
<\Omega, \tilde{u}(x_{1}) u(x_{2})\Omega> = i~D_{m}^{(+)}(x_{1} - x_{2})
\eea
and the $n$-point functions are generated according to Wick theorem. Here
$
D_{m}^{(+)}
$
is the positive frequency part of the Pauli-Jordan distribution
$
D_{m}
$
of mass $m$. To extend the sesquilinear form to
$
{\cal H}
$
we define the conjugation by
\be
v_{\mu}^{\dagger} = v_{\mu}, \qquad 
u^{\dagger} = u, \qquad
\tilde{u}^{\dagger} = - \tilde{u},
\qquad \Phi^{\dagger} = \Phi.
\ee

Now we can define in 
$
{\cal H}
$
the operator $Q$ according to the following formulas:
\bea
~[Q, v_{\mu}] = i~\partial_{\mu}u,\qquad
[Q, u] = 0,\qquad
[Q, \tilde{u}] = - i~(\partial_{\mu}v^{\mu} + m~\Phi)
\qquad
[Q,\Phi] = i~m~u,
\nonumber \\
Q\Omega = 0.
\label{Q-m}
\eea
One can prove that $Q$ is well defined. We have a result similar to the first
theorem of this Section:
\begin{thm}
The operator $Q$ verifies
$
Q^{2} = 0.
$ 
The factor space
$
Ker(Q)/Ran(Q)
$
is isomorphic to the Fock space of particles of mass $m$ and spin $1$ (massive
photons). 
\end{thm}
%\newpage
\subsection{The Generic Yang-Mills Case}

The situations described above (of massless and massive photons) are susceptible
of the following generalizations. We can consider a system of 
$
r_{1}
$ 
species of particles of null mass and helicity $1$ if we use in the first part
of this Section 
$
r_{1}
$ 
triplets
$
(v^{\mu}_{a}, u_{a}, \tilde{u}_{a}), a \in I_{1}
$
of massless fields; here
$
I_{1}
$
is a set of indexes of cardinal 
$
r_{1}.
$
All the relations have to be modified by appending an index $a$ to all these
fields. 

In the massive case we have to consider 
$
r_{2}
$ 
quadruples
$
(v^{\mu}_{a}, u_{a}, \tilde{u}_{a}, \Phi_{a}),  a \in I_{2}
$
of fields of mass 
$
m_{a}
$; here
$
I_{2}
$
is a set of indexes of cardinal 
$
r_{2}.
$

We can consider now the most general case involving fields of spin not greater
that $1$.
We take 
$
I = I_{1} \cup I_{2} \cup I_{3}
$
a set of indexes and for any index we take a quadruple
$
(v^{\mu}_{a}, u_{a}, \tilde{u}_{a},\Phi_{a}), a \in I
$
of fields with the following conventions:
(a) For
$
a \in I_{1}
$
we impose 
$
\Phi_{a} = 0
$
and we take the masses to be null
$
m_{a} = 0;
$
(b) For
$
a \in I_{2}
$
we take the all the masses strictly positive:
$
m_{a} > 0;
$
(c) For 
$
a \in I_{3}
$
we take 
$
v_{a}^{\mu}, u_{a}, \tilde{u}_{a}
$
to be null and the fields
$
\Phi_{a} \equiv \phi^{H}_{a} 
$
of mass 
$
m^{H}_{a} \geq 0.
$
The fields
$
\phi^{H}_{a} 
$
are called {\it Higgs fields}.

If we define
$
m_{a} = 0, \forall a \in I_{3}
$
then we can define in 
$
{\cal H}
$
the operator $Q$ according to the following formulas for all indexes
$
a \in I:
$
\bea
~[Q, v^{\mu}_{a}] = i~\partial^{\mu}u_{a},\qquad
[Q, u_{a}] = 0,
\nonumber \\
~[Q, \tilde{u}_{a}] = - i~(\partial_{\mu}v^{\mu}_{a} + m_{a}~\Phi_{a})
\qquad
[Q,\Phi_{a}] = i~m_{a}~u_{a},
\nonumber \\
Q\Omega = 0.
\label{Q-general}
\eea

If we consider matter fields also i.e some set of Dirac fields with Fermi
statistics:
$
\psi_{A}, A \in I_{4}
$ 
then we impose
\be
d_{Q}\psi_{A} = 0.
\ee

\newpage
\subsection{The Yang-Mills Interaction\label{interaction}}

In the framework and notations from the end of the preceding Section we have the 
following result which describes the most general form of the Yang-Mills interaction
\cite{YM}, \cite{standard}, \cite{fermi}. Summation over the dummy indexes is used 
everywhere.
\begin{thm}
Let $T$ be a relative cocycle for 
$
d_{Q}
$
which is as least tri-linear in the fields and is of canonical dimension
$
\omega(T) \leq 4
$
and ghost number
$
gh(T) = 0.
$
Then:
(i) $T$ is (relatively) cohomologous to a non-trivial co-cycle of the form:
\bea
T = f_{abc} \left( {1\over 2}~v_{a\mu}~v_{b\nu}~F_{c}^{\nu\mu}
+ u_{a}~v_{b}^{\mu}~\partial_{\mu}\tilde{u}_{c}\right)
\nonumber \\
+ f^{\prime}_{abc} (\Phi_{a}~\phi_{b}^{\mu}~v_{c\mu} +
m_{b}~\Phi_{a}~\tilde{u}_{b}~u_{c})
\nonumber \\
+ {1\over 3!}~f^{\prime\prime}_{abc}~\Phi_{a}~\Phi_{b}~\Phi_{c}
+ j^{\mu}_{a}~v_{a\mu} + j_{a}~\Phi_{a};
\eea
where we can take the constants
$
f_{abc} = 0
$
if one of the indexes is in
$
I_{3};
$
also
$
f^{\prime}_{abc} = 0
$
if 
$
c \in I_{3}
$
or one of the indexes $a$ and $b$ are from
$
I_{1};
$
and
$
j^{\mu}_{a} = 0
$
if
$
a \in I_{3};
$
$
j_{a} = 0
$
if
$
a \in I_{1}.
$
By definition
\be
\phi_{a}^{\mu} \equiv \partial^{\mu}\Phi_{a} - v_{a}^{\mu}
\ee
Moreover we have:

(a) The constants
$
f_{abc}
$
are completely antisymmetric
\be
f_{abc} = f_{[abc]}.
\label{anti-f}
\ee

(b) The expressions
$
f^{\prime}_{abc}
$
are antisymmetric  in the indexes $a$ and $b$:
\be
f^{\prime}_{abc} = - f^{\prime}_{bac}
\label{anti-f'}
\ee
and are connected to 
$f_{abc}$
by:
\be
f_{abc}~m_{c} = f^{\prime}_{cab} m_{a} - f^{\prime}_{cba} m_{b}.
\label{f-f'}
\ee

(c) The (completely symmetric) expressions 
$f^{\prime\prime}_{abc} = f^{\prime\prime}_{\{abc\}}$
verify
\be
f^{\prime\prime}_{abc}~m_{c} = \left\{\begin{array}{rcl} 
{1 \over m_{c}}~f'_{abc}~(m_{a}^{2} - m_{b}^{2}) & \mbox{for} & a, b \in I_{3},
c \in I_{2} \\
- {1 \over m_{c}}~f'_{abc}~m_{b}^{2} & \mbox{for} & a, c \in I_{2}, b \in
I_{3}.\end{array}\right.
\label{f"}
\ee

(d) the expressions
$
j^{\mu}_{a}
$
and
$
j_{a}
$
are bilinear in the Fermi matter fields: in tensor notations;
\bea
j_{a}^{\mu} = \sum_{\epsilon}~
\overline{\psi} t^{\epsilon}_{a} \otimes \gamma^{\mu}\gamma_{\epsilon} \psi
\qquad
%\nonumber \\
j_{a} = \sum_{\epsilon}~
\overline{\psi} s^{\epsilon}_{a} \otimes \gamma_{\epsilon} \psi
\label{current}
\eea
where  for every
$
\epsilon = \pm
$
we have defined the chiral projectors of the algebra of Dirac matrices
$
\gamma_{\epsilon} \equiv {1\over 2}~(I + \epsilon~\gamma_{5})
$
and
$
t^{\epsilon}_{a},~s^{\epsilon}_{a}
$
are 
$
|I_{4}| \times |I_{4}|
$
matrices. If $M$ is the mass matrix
$
M_{AB} = \delta_{AB}~M_{A}
$
then we must have
\be
\partial_{\mu}j^{\mu}_{a} = m_{a}~j_{a} 
\qquad \Leftrightarrow \qquad
m_{a}~s_{a}^{\epsilon} = i(M~t^{\epsilon}_{a} - t^{-\epsilon}_{a}~M).
\label{conserved-current}
\ee

(ii) The relation 
$
d_{Q}T = i~\partial_{\mu}T^{\mu}
$
is verified by:
\be
T^{\mu} = f_{abc} \left( u_{a}~v_{b\nu}~F^{\nu\mu}_{c} -
{1\over 2} u_{a}~u_{b}~\partial^{\mu}\tilde{u}_{c} \right)
+ f^{\prime}_{abc}~\Phi_{a}~\phi_{b}^{\mu}~u_{c}
+ j^{\mu}_{a}~u_{a}
\label{Tmu}
\ee

(iii) The relation 
$
d_{Q}T^{\mu} = i~\partial_{\nu}T^{\mu\nu}
$
is verified by:
\be
T^{\mu\nu} \equiv {1\over 2} f_{abc}~u_{a}~u_{b}~F_{c}^{\mu\nu}.
\ee
\label{T1}
\end{thm}
\newpage
\section{Causal Perturbation Theory\label{causal}}
 
We give here the essential ingredients of perturbation theory. We consider that
the canonical dimension of the vector and scalar fields
$
v_{a}^{\mu}, u_{a}, \tilde{u}_{a}, \Phi_{a}
$
is equal to $1$ and the canonical dimension of the Dirac fields is
$
3/2
$.
A derivative applied to a field raises the canonical dimension by $1$. The ghost
number of the ghost fields is $1$ and for the rest of the fields is null. The
Fermi parity of a Fermi (Bose) field is $1$ (resp. $0$). The canonical 
dimension of a Wick monomial is additive with respect to the factors and the
same is true for the ghost number and the Fermi parity.

\subsection{Bogoliubov Axioms}{\label{bogoliubov}}

Suppose that the Wick monomials
$
W_{1},\dots,W_{n}
$
are self-adjoint:
$
W_{j}^{\dagger} = W_{j},~\forall j = 1,\dots,n.
$
The chronological products
$ 
T(W_{1}(x_{1}),\dots,W_{n}(x_{n})) \quad n = 1,2,\dots
$
are verifying the following set of axioms:
\begin{itemize}
\item
Skew-symmetry in all arguments
$
W_{1}(x_{1}),\dots,W_{n}(x_{n}):
$
\be
T(\dots,W_{i}(x_{i}),W_{i+1}(x_{i+1}),\dots,) =
(-1)^{f_{i} f_{i+1}} T(\dots,W_{i+1}(x_{i+1}),W_{i}(x_{i}),\dots)
\ee
where
$f_{i}$
is the number of Fermi fields appearing in the Wick monomial
$W_{i}$.
\item
Poincar\'e invariance: we have a natural action of the Poincar\'e group in the
space of Wick monomials and we impose that for all 
$(a,A) \in inSL(2,\C)$
we have:
\be
U_{a, A} T(W_{1}(x_{1}),\dots,W_{n}(x_{n})) U^{-1}_{a, A} =
T(A\cdot W_{1}(A\cdot x_{1}+a),\dots,A\cdot W_{n}(A\cdot x_{n}+a));
\label{invariance}
\ee

Sometimes it is possible to supplement this axiom by other invariance
properties: space and/or time inversion, charge conjugation invariance, global
symmetry invariance with respect to some internal symmetry group, supersymmetry,
etc.
\item
Causality: if
$x_{i} \geq x_{j}, \quad \forall i \leq k, \quad j \geq k+1$
then we have:
\be
T(W_{1}(x_{1}),\dots,W_{n}(x_{n})) =
T(W_{1}(x_{1}),\dots,W_{k}(x_{k}))~~T(W_{k+1}(x_{k+1}),\dots,W_{n}(x_{n}));
\label{causality}
\ee
\item
Unitarity: We define the {\it anti-chronological products} according to
\be
(-1)^{n} \bar{T}(W_{1}(x_{1}),\dots,W_{n}(x_{n})) \equiv \sum_{r=1}^{n} 
(-1)^{r} \sum_{I_{1},\dots,I_{r} \in Part(\{1,\dots,n\})}
\epsilon~~T_{I_{1}}(X_{1})\cdots T_{I_{r}}(X_{r})
\label{antichrono}
\ee
where the we have used the notation:
\be
T_{\{i_{1},\dots,i_{k}\}}(x_{i_{1}},\dots,x_{i_{k}}) \equiv 
T(W_{i_{1}}(x_{i_{1}}),\dots,W_{i_{k}}(x_{i_{k}}))
\ee
and the sign
$\epsilon$
counts the permutations of the Fermi factors. Then the unitarity axiom is:
\be
\bar{T}(W_{1}(x_{1}),\dots,W_{n}(x_{n})) =
T(W_{1}(x_{1}),\dots,W_{n}(x_{n}))^{\dagger}.
\label{unitarity}
\ee
\item
The ``initial condition"
\be
T(W(x)) = W(x).
\ee
\end{itemize}

It can be proved that this system of axioms can be supplemented with
\bea
T(W_{1}(x_{1}),\dots,W_{n}(x_{n}))
\nonumber \\
= \sum \quad
<\Omega, T(W^{\prime}_{1}(x_{1}),\dots,W^{\prime}_{n}(x_{n}))\Omega>~~
:W^{\prime\prime}_{1}(x_{1}),\dots,W^{\prime\prime}_{n}(x_{n}):
\label{wick-chrono2}
\eea
where
$W^{\prime}_{i}$
and
$W^{\prime\prime}_{i}$
are Wick submonomials of
$W_{i}$
such that
$W_{i} = :W^{\prime}_{i} W^{\prime\prime}_{i}:$
and we have supposed that only Bose fields are present; if Fermi fields are
present then some apropriate signs should be inserted.
This is called the {\it Wick expansion property}. 

We can also include in the induction hypothesis a limitation on the order of
singularity of the vacuum averages of the chronological products associated to
arbitrary Wick monomials
$W_{1},\dots,W_{n}$;
explicitly:
\be
\omega(<\Omega, T^{W_{1},\dots,W_{n}}(X)\Omega>) \leq
\sum_{l=1}^{n} \omega(W_{l}) - 4(n-1)
\label{power}
\ee
where by
$\omega(d)$
we mean the order of singularity of the (numerical) distribution $d$ and by
$\omega(W)$
we mean the canonical dimension of the Wick monomial $W$; in particular this
means
that we have
\be
T(W_{1}(x_{1}),\dots,W_{n}(x_{n}))
= \sum_{g} t_{g}(x_{1},\dots,x_{n})~W_{g}(x_{1},\dots,x_{n})
\label{generic}
\ee
where
$W_{g}$
are Wick polynomials of fixed canonical dimension and
$t_{g}$
are distributions in 
$
n - 1
$
variables (because of translation invariance) with the order of singularity
bounded by the power counting
theorem \cite{EG}:
\be
\omega(t_{g}) + \omega(W_{g}) \leq
\sum_{j=1}^{n} \omega(W_{j}) - 4 (n - 1)
\label{power1}
\ee
and the sum over $g$ is essentially a sum over Feynman graphs. Up to now, we
have defined the chronological products only for self-adjoint Wick monomials 
$
W_{1},\dots,W_{n}
$
but we can extend the definition for Wick polynomials by linearity.

The basic construction of Epstein and Glaser is the construction of the causal
commutator. In the second order of the perturbation theory this is simply
\be
D(A(x),B(y)) = [ A(x),B(y)]
\label{com2}
\ee
where 
$
A(x), B(y)
$
are arbitrary Wick monomials and $
[\cdot,\cdot]
$
the graded commutator. This distribution is translation invariant and with causal
support i.e. it depends only on 
$
x - y
$
and the support is inside the light cones:
\be
supp(D) \subset V^{+} \cup V^{-}.
\ee
The simple formula (\ref{com2}) and support property is the justification of the 
terminology of {\it causal commutator}. 

In higher orders of the perturbation theory the generalization of (\ref{com2}) is more
complicated but we need only the third-order formula which is for even
$
A, B, C
$:
\bea
D(A(x),B(y),C(z)) =
\nonumber\\
- [\bar{T}(A(x),B(y)),C(z)] - [T(A(x),C(z)), B(y)] - [T(B(y),C(z)), A(y)]
\label{comm3}
\eea
and in the general case all commutators are graded and we insert appropriate signes. 
Again one can prove that this distribution is translation invariant and with causal
support i.e. it depends only on the variables
$
x - z, y - z
$
and has the support in the causal cone
\be
V^{\rm causal} \equiv \{ (x,y,z) | x - z \in V^{+}, y - z \in V^{+} \} \cup 
\{(x,y,z) | x - z \in V^{-}, y - z \in V^{-}  \}.
\label{causal-cone}
\ee
%\newpage
\subsection{Third Order Causal Distributions}

We remind the fact that the Pauli-Villars distribution is defined by
\be
D_{m}(x) = D_{m}^{(+)}(x) + D_{m}^{(-)}(x)
\ee
where 
\be
D_{m}^{(\pm)}(x) \sim 
\int dp e^{i p\cdot x} \theta(\pm p_{0}) \delta(p^{2} - m^{2})
\ee
such that
\be
D^{(-)}(x) = - D^{(+)}(- x).
\ee

This distribution has causal support. In fact, it can be causally split into an
advanced and a retarded part:
\be
D = D^{\rm adv} - D^{\rm ret}
\ee
and then we can define the Feynman propagator and antipropagator
\be
D^{F} = D^{\rm ret} + D^{(+)}, \qquad \bar{D}^{F} = D^{(+)} - D^{\rm adv}.
\ee
All these distributions have singularity order
$
\omega(D) = -2
$.

For the triangle loop contributions in the third order we need some basic
distributions. First, we take
$
D_{j} = D_{m_{j}}, j = 1,2,3
$
and define
\bea
d_{D_{1},D_{2},D_{3}}(x,y,z) \equiv \bar{D}^{F}_{3}(x - y) 
[ D^{(-)}_{2}(z - x) D^{(+)}_{1}(y - z) - D^{(+)}_{2}(z - x) D^{(-)}_{1}(y - z)
]
\nonumber \\
+ D^{F}_{1}(y - z) 
[ D^{(-)}_{3}(x - y) D^{(+)}_{2}(z - x) - D^{(+)}_{3}(x - y) D^{(-)}_{2}(z - x)
]
\nonumber \\
+ D^{F}_{2}(z - x) 
[ D^{(-)}_{1}(y - z) D^{(+)}_{3}(x - y) - D^{(+)}_{1}(y - z) D^{(-)}_{3}(x - y)
]
\eea
which also with causal support; indeed we have the alternative forms
\bea
d_{D_{1},D_{2},D_{3}}(x,y,z) = - D^{\rm ret}_{3}(x - y) 
[ D^{(-)}_{2}(z - x) D^{(+)}_{1}(y - z) - D^{(+)}_{2}(z - x) D^{(-)}_{1}(y - z)
]
\nonumber \\
+ D^{\rm adv}_{1}(y - z) 
[ D^{(-)}_{3}(x - y) D^{(+)}_{2}(z - x) - D^{(+)}_{3}(x - y) D^{(-)}_{2}(z - x)
]
\nonumber \\
+ D^{\rm adv}_{2}(z - x) 
[ D^{(-)}_{1}(y - z) D^{(+)}_{3}(x - y) - D^{(+)}_{1}(y - z) D^{(-)}_{3}(x - y)
]
\eea
and
\bea
d_{D_{1},D_{2},D_{3}}(x,y,z) = - D^{\rm adv}_{3}(x - y) 
[ D^{(-)}_{2}(z - x) D^{(+)}_{1}(y - z) - D^{(+)}_{2}(z - x) D^{(-)}_{1}(y - z)
]
\nonumber \\
+ D^{\rm ret}_{1}(y - z) 
[ D^{(-)}_{3}(x - y) D^{(+)}_{2}(z - x) - D^{(+)}_{3}(x - y) D^{(-)}_{2}(z - x)
]
\nonumber \\
+ D^{\rm ret}_{2}(z - x) 
[ D^{(-)}_{1}(y - z) D^{(+)}_{3}(x - y) - D^{(+)}_{1}(y - z) D^{(-)}_{3}(x - y)
]
\eea
from which it follows that 
$
d_{D_{1},D_{2},D_{3}}(x,y,z)
$
is null outside the causal cone (\ref{causal-cone}). These distributions have the 
singularity order
$
\omega(d_{D_{1},D_{2},D_{3}}) = - 2
$.

There are some associated distributions obtained from
$
d_{D_{1},D_{2},D_{3}}(x,y,z)
$
applying derivatives on the factors
$
D_{j} = D_{m_{j}}, j = 1,2,3
$.
For instance we denote
\bea
{\cal D}^{1}_{\alpha}d_{D_{1},D_{2},D_{3}} \equiv
d_{\partial_{\alpha}D_{1},D_{2},D_{3}}
\nonumber \\
{\cal D}^{2}_{\alpha}d_{D_{1},D_{2},D_{3}} \equiv
d_{D_{1},\partial_{\alpha}D_{2},D_{3}}
\nonumber \\
{\cal D}^{3}_{\alpha}d_{D_{1},D_{2},D_{3}} \equiv
d_{D_{1},D_{2},\partial_{\alpha}D_{3}},
\eea
and so on for more derivatives
$
\partial_{\alpha}
$
distributed in an arbitrary way on the factors
$
D_{j} = D_{m_{j}}, j = 1,2,3
$.
We mention the fact that the operators
$
{\cal D}^{j}_{\alpha}, j = 1,2,3
$
are commutative but they are not derivation operators: they do not verify
Leibnitz rule.

When it possible we skip the dependence on
$
D_{j} = D_{m_{j}}, j = 1,2,3
$
i.e. we simply write
$
d = d_{D_{1},D_{2},D_{3}}.
$
We note the formulas
\bea
{\partial \over \partial x^{\alpha}}d = 
( {\cal D}^{3}_{\alpha} - {\cal D}^{2}_{\alpha})d
\nonumber \\
{\partial \over \partial y^{\alpha}}d =
( {\cal D}^{1}_{\alpha} - {\cal D}^{3}_{\alpha})d
\nonumber \\
{\partial \over \partial z^{\alpha}}d =
( {\cal D}^{2}_{\alpha} - {\cal D}^{1}_{\alpha})d
\eea

Apparently, these distributions do not have nice symmetry properties in all the
three variables. However, this is not true. Let
$
A(x), B(y), C(z)
$ 
be three Wick monomials. Then the triangle one-loop contribution of the causal
commutator is of the form:
\be
D_{\rm triangle}(A(x), B(y), C(z)) = \sum 
p_{j}({\cal D}^{1},{\cal D}^{2},{\cal D}^{3})d_{j}(x,y,z)~W_{j}(x,y,z)
\label{dABC1}
\ee
where
$
d_{j}
$
are distributions of the type
$
d_{D_{1},D_{2},D_{3}}
$
$,
p_{j}
$
are polynomials in the operators
$
{\cal D}^{j}_{\alpha}, j = 1,2,3
$
and 
$
W_{j}(x,y,z)
$
are Wick monomials. For simplicity, let us suppose that the monomials
$
A(x), B(y), C(z)
$ 
are even so they causally commute. Then we have
\bea
D_{\rm triangle}(B(x), A(y), C(z)) = \sum 
p_{j}(- {\cal D}^{2}, - {\cal D}^{1},- {\cal D}^{3})d_{j}(x,y,z)~W_{j}(y,x,z)
\nonumber \\
D_{\rm triangle}(A(x), C(y), B(z)) = \sum 
p_{j}(- {\cal D}^{1}, - {\cal D}^{3},- {\cal D}^{2})d_{j}(x,y,z)~W_{j}(x,z,y)
\nonumber \\
D_{\rm triangle}(C(x), B(y), A(z)) = \sum 
p_{j}(- {\cal D}^{3}, - {\cal D}^{2},- {\cal D}^{1})d_{j}(x,y,z)~W_{j}(z,y,x)
\label{dABC2}
\eea
i.e. the exhange of the factors
$
A, B, C
$ 
can be accounted for in a natural way. If some of the monomials
$
A(x), B(y), C(z)
$ 
are odd, the some signes must be inserted in the preceding sums. For
instance, is $A$ and $B$ are causally anticommuting, then we have an extra $-$
sign in the first and the third line above.
\newpage
In the third order of perturbation theory other causal distributions can appear.
They are associated with the so-called {\it one-particle reducible Feynman graphs}.
\bea
d^{(1)}_{D_{1},D_{2}}(x,y,z) \equiv 
\bar{D}^{F}_{1}(x - y) D_{2}(z - x) - D_{1}(x - y) D^{F}_{2}(z - x)
\nonumber \\
+ D^{(-)}_{1}(x - y) D^{(+)}_{2}(z - x) - D^{(+)}_{1}(x - y) D^{(-)}_{2}(z - x)
]
\nonumber \\
d^{(2)}_{D_{1},D_{2}}(x,y,z) \equiv 
- \bar{D}^{F}_{1}(x - y) D_{2}(y - z) + D_{1}(x - y) D^{F}_{2}(y - z)
\nonumber \\
+ D^{(+)}_{1}(x - y) D^{(-)}_{2}(y - z) - D^{(-)}_{1}(x - y) D^{(+)}_{2}(y - z)
]
\nonumber \\
d^{(3)}_{D_{1},D_{2}}(x,y,z) \equiv 
D^{F}_{1}(z - x) D_{2}(y - z) - D_{1}(z - x) D^{F}_{2}(y - z)
\nonumber \\
+ D^{(-)}_{1}(z - x) D^{(+)}_{2}(y - z) - D^{(+)}_{1}(z - x) D^{(-)}_{2}(y - z)
]
\eea
The causal support properties follow from the alternative formulas
\bea
d^{(1)}_{D_{1},D_{2}}(x,y,z) =
D^{\rm ret}_{1}(x - y) D^{\rm ret}_{2}(z - x) 
- D^{\rm adv}_{1}(x - y) D^{\rm adv}_{2}(z - x)
\nonumber \\
d^{(2)}_{D_{1},D_{2}}(x,y,z) = D^{\rm ret}_{1}(y - x) D^{\rm ret}_{2}(z - y) 
- D^{\rm adv}_{1}(y - x) D^{\rm adv}_{2}(z - y)
\nonumber \\
d^{(3)}_{D_{1},D_{2}}(x,y,z) = D^{\rm ret}_{1}(z - x) D^{\rm ret}_{2}(y - z) 
- D^{\rm adv}_{1}(z - x) D^{\rm adv}_{2}(y - z)
\eea
and this leads to the following Feynman propagators
\bea
d^{(1)F}_{D_{1},D_{2}}(x,y,z) = D^{F}_{1}(x - y) D^{F}_{2}(z - x)
\nonumber \\
d^{(2)F}_{D_{1},D_{2}}(x,y,z) = D^{F}_{1}(y - x) D^{F}_{2}(z - y)
\nonumber \\
d^{(3)F}_{D_{1},D_{2}}(x,y,z) = D^{F}_{1}(z - x) D^{F}_{2}(y - z)
\eea

The order of singularity of these distributions is again
$
\omega = - 2
$.
We can define associated distributions as before if we replace
$
D_{1} \mapsto \partial_{\alpha}D_{1}
$,
etc. 
We need to consider the case when one of the distribution
$
D_{1}, D_{2}
$
is of the type
$
D_{m}
$
and the other is of the type
$
d_{2}
$
where 
\be
d_{2}(x) \equiv {1\over 2} [ D_{m}^{(+)}(x)^{2} - D_{m}^{(+)}(- x)^{2} ]
\ee
Let us notice that some associated distributions can have some $\delta$ factors.
We denote
\be
{\cal K}_{j} = {\cal D}_{j}^{\mu} {\cal D}_{j\mu},~j = 1,2,3
\ee
and we have for instance
\bea
{\cal K}_{1}d(x,y,z) = 2 \delta(y - z) d_{2}(x - y)
\nonumber\\
{\cal K}_{2}d(x,y,z) = 2 \delta(z - x) d_{2}(y - z)
\nonumber\\
{\cal K}_{3}d(x,y,z) = - 2 \delta(x - y) d_{2}(z - x)
\label{on-shell}
\eea
and similar relations for the distributions
$
d^{(j)}~,j = 1,2,3
$.

\newpage
\section{Super-Renormalizablity in the Third Order\label{sr}}
We need the explicit form of the causal commutators
$
D^{IJK}
$.
From (\ref{dABC1}) and (\ref{dABC2}) we can obtain some symmetry properties. We
also have the ghost number restrictions
\be
gh(D^{IJK}) = |I| + |J| + |K|.
\label{ghD}
\ee
If we use for them the generic form (\ref{generic}) then (\ref{power1}) gets the 
form
\be
\omega(t_{g}) + \omega(W_{g}) \leq 4.
\label{powerD}
\ee
We suppose that we have establish gauge invariance up to the second order of the
perturbation theory i.e.
\be
(sT)^{I}(x) = 0, (sT)^{IJ}(x,y) = 0
\ee
and we obtain from the definition the cocycle property
\be
(sD)^{IJK}(x,y,z) = 0.
\ee
We determine under what conditions the expression
$
D^{IJK}
$
are coboundaries, up to super-renormalizable terms i.e.
\be
D^{IJK}(x,y,z) = (\bar{s}B)^{IJK}(x,y,z) + {\rm super-renormalizable~terms}
\label{co}
\ee
and we will need the generic form for the coboundaries
$
B^{IJK}
$.

Like in \cite{super2} we replace everywhere for every mass $m$ in the game
\be
D_{m} = D_{0} + ( D_{m} - D_{0})
\label{split}
\ee
In this way we split 
$
D_{(1)}^{IJK}(x,y,z)
$
into a contribution 
$
D_{(1)0}^{IJK}(x,y,z)
$
where everywhere
$
D_{m} \mapsto D_{0}
$
and a contribution where at least one factor
$
D_{m}
$
is replaced by the difference
$
D_{m} - D_{0}
$.
Because we have
\be
\omega(D_{m} - D_{0}) = -4
\ee
the second contribution will be super-renormalizable. We need to consider only the first
contribution 
$
D_{(1)0}^{IJK}(x,y,z)
$
and investigate if it can be written as coboundary. In the preceding expression 
we have two type of terms: ones associated to the triangle graphs and the other 
associated to the one-particle reducible graphs.
\be
D_{(1)0}^{IJK}(x,y,z) = D_{\rm triangle}^{IJK}(x,y,z) + D_{\rm 1PR}^{IJK}(x,y,z)
\label{graphsplit}
\ee

From (\ref{dABC1}) and (\ref{dABC2}) we can obtain that the triangle
contribution
$
D^{IJK}_{\rm triangle}(x,y,z)
$
is invariant with respect to the following transformation, up to some signs;
for instance if $I,J,K$ are even, we have invariance with respect to
\bea
x \leftrightarrow y, I \leftrightarrow J, 
{\cal D}_{1} \rightarrow - {\cal D}_{2}, 
{\cal D}_{2} \rightarrow - {\cal D}_{1},
{\cal D}_{3} \rightarrow - {\cal D}_{3}
\nonumber \\
y \leftrightarrow z, J \leftrightarrow K, 
{\cal D}_{1} \rightarrow - {\cal D}_{1}, 
{\cal D}_{2} \rightarrow - {\cal D}_{3},
{\cal D}_{3} \rightarrow - {\cal D}_{2}
\nonumber \\
x \leftrightarrow z, I \leftrightarrow K, 
{\cal D}_{1} \rightarrow - {\cal D}_{3}, 
{\cal D}_{2} \rightarrow - {\cal D}_{2},
{\cal D}_{3} \rightarrow - {\cal D}_{2}
\label{sD}
\eea
and in the general case we have invariance up to the sign
$
(-1)^{|I||J|}
$
in the first case, etc. Similar symmetry properties are valid for the
one-particle irreducible contribution.

In both terms from the righthand side of (\ref{graphsplit}) we have
delta-contributions i.e. contributions where the operators
$
{\cal K}_{j}~j = 1,2,3
$
are present and non-delta-contributions i.e. contributions where the  operators
$
{\cal K}_{j}~j = 1,2,3
$
are absent; this splitting is unique.
\bea
D_{\rm triangle}^{IJK}(x,y,z) = 
D_{\rm triangle}^{IJK}(x,y,z)_{0} + D_{\rm triangle}^{IJK}(x,y,z)_{\delta}
\nonumber\\
D_{\rm 1PR}^{IJK}(x,y,z) = 
D_{\rm 1PR}^{IJK}(x,y,z)_{0} + D_{\rm 1PR}^{IJK}(x,y,z)_{\delta}
\eea
If
$
D_{(1)0}^{IJK}(x,y,z)
$
is a coboundary, then the expressions
$
D_{\rm triangle}^{IJK}(x,y,z)_{0}
$
and
$
D_{\rm 1PR}^{IJK}(x,y,z)_{0}
$
should also be coboundaries, up to delta-contributions. If this is establish we still 
have to check that the sum of the delta-contributions from the triangle and the
1PR contributions can be also written as a coboundary.

We have to compute explicitly the contributions
$
D_{\rm triangle}^{IJK}(x,y,z)_{0}
$
and
$
D_{\rm 1PR}^{IJK}(x,y,z)_{0}
$.
It is easy to start a descent procedure and to consider the case of maximal ghost number
$
|I| + |J| + |K| = 3
$.
It is useful to give the generic form compatible with the symmetry property in all
three variables (given in the preceding Section) for
$
D^{IJK}_{\rm triangle},~|I| + |J| + |K| = 3
$. 
Having the generic form one can prove that the cohomology is not trivial i.e. from the 
cocycle identity we cannot obtain the coboundary property. So in the end we will have to 
compute explicitly the preceding commutator. The computation are straightforward but
rather long so we will present only some of them.

We consider the equation
\be
D_{\rm triangle}^{IJK}(x,y,z)_{0} = (\bar{s}B)^{IJK}(x,y,z) + D_{\delta}^{IJK}(x,y,z)
\ee
where the expressions
$
B^{IJK}
$
are restricted by  skew-symmetry properties of the type (\ref{sD}), ghost number
restrictions
\be
gh(B^{IJK}(x,y,z)) = |I| + |J| + |K| - 1
\label{ghB}
\ee
and power counting
\be
\omega(t_{g}) + \omega(W_{g}) \leq 3
\label{powerB}
\ee
similar to (\ref{ghD}) and (\ref{powerD}) respectively.
\newpage
\subsection{The Causal Commutators of Ghost Number $3$}
We first consider the causal commutator
\bea
D^{[\mu][\nu][\rho]}(x,y,z) = 
D(T^{\mu}(x),T^{\nu}(y);T^{\rho}(z))
\nonumber \\
- [ \bar{T}(T^{\mu}(x),T^{\nu}(y)),T^{\rho}(z) ]
+ [ T(T^{\mu}(x),T^{\rho}(z)),T^{\nu}(y) ]
- [ T(T^{\nu}(y),T^{\rho}(z)),T^{\mu}(x) ].
\eea
The generic form of the triangle contribution is:
\bea
D_{\rm triangle}^{[\mu],[\nu],[\rho]}(x,y,z)_{0} = i
[ 2 A_{1, abc} ({\cal D}_{1}^{\mu}{\cal D}_{1}^{\nu}{\cal D}_{1}^{\rho}
+ {\cal D}_{2}^{\mu}{\cal D}_{2}^{\nu}{\cal D}_{2}^{\rho}
+ {\cal D}_{3}^{\mu}{\cal D}_{3}^{\nu}{\cal D}_{3}^{\rho})
\nonumber \\
+ A_{2, abc} ({\cal D}_{1}^{\mu}{\cal D}_{1}^{\nu}{\cal D}_{2}^{\rho}
+ {\cal D}_{1}^{\rho}{\cal D}_{2}^{\mu}{\cal D}_{2}^{\nu}
+ {\cal D}_{1}^{\mu}{\cal D}_{1}^{\rho}{\cal D}_{3}^{\nu}
+ {\cal D}_{1}^{\nu}{\cal D}_{3}^{\mu}{\cal D}_{3}^{\rho}
+ {\cal D}_{2}^{\nu}{\cal D}_{2}^{\rho}{\cal D}_{3}^{\mu}
+ {\cal D}_{2}^{\mu}{\cal D}_{3}^{\nu}{\cal D}_{3}^{\rho})
\nonumber \\
+ A_{3, abc} ({\cal D}_{1}^{\mu}{\cal D}_{1}^{\rho}{\cal D}_{2}^{\nu}
+ {\cal D}_{1}^{\mu}{\cal D}_{2}^{\nu}{\cal D}_{2}^{\rho}
+ {\cal D}_{1}^{\mu}{\cal D}_{1}^{\nu}{\cal D}_{3}^{\rho}
+ {\cal D}_{1}^{\mu}{\cal D}_{3}^{\nu}{\cal D}_{3}^{\rho}
+ {\cal D}_{2}^{\mu}{\cal D}_{2}^{\nu}{\cal D}_{3}^{\rho}
+ {\cal D}_{2}^{\mu}{\cal D}_{3}^{\nu}{\cal D}_{3}^{\rho})
\nonumber \\
+ A_{4, abc} ({\cal D}_{1}^{\nu}{\cal D}_{1}^{\rho}{\cal D}_{2}^{\mu}
+ {\cal D}_{1}^{\nu}{\cal D}_{2}^{\mu}{\cal D}_{2}^{\rho}
+ {\cal D}_{1}^{\nu}{\cal D}_{1}^{\rho}{\cal D}_{3}^{\mu}
+ {\cal D}_{1}^{\rho}{\cal D}_{3}^{\mu}{\cal D}_{3}^{\nu}
+ {\cal D}_{2}^{\mu}{\cal D}_{2}^{\rho}{\cal D}_{3}^{\nu}
+ {\cal D}_{2}^{\rho}{\cal D}_{3}^{\mu}{\cal D}_{3}^{\nu})
\nonumber \\
+ 6 A_{5, abc} {\cal D}_{1}^{\mu}{\cal D}_{2}^{\nu}{\cal D}_{3}^{\rho}
\nonumber \\
+ 2 A_{6, abc} ({\cal D}_{1}^{\mu}{\cal D}_{2}^{\rho}{\cal D}_{3}^{\nu}
+ {\cal D}_{1}^{\rho}{\cal D}_{2}^{\nu}{\cal D}_{3}^{\mu}
+ {\cal D}_{1}^{\nu}{\cal D}_{2}^{\mu}{\cal D}_{3}^{\rho})
\nonumber \\
+ 3 A_{7, abc} ({\cal D}_{1}^{\rho}{\cal D}_{2}^{\mu}{\cal D}_{3}^{\nu}
+ {\cal D}_{1}^{\nu}{\cal D}_{2}^{\rho}{\cal D}_{3}^{\mu})
\nonumber \\
+ A_{8, abc} (\eta^{\mu\nu} {\cal D}_{1}^{\rho}{\cal D}_{1}\cdot{\cal D}_{2}
+ \eta^{\mu\nu} {\cal D}_{2}^{\rho}{\cal D}_{1}\cdot{\cal D}_{2}
+ \eta^{\mu\rho} {\cal D}_{1}^{\nu}{\cal D}_{1}\cdot{\cal D}_{3}
\nonumber \\
+ \eta^{\mu\rho} {\cal D}_{3}^{\nu}{\cal D}_{1}\cdot{\cal D}_{3}
+ \eta^{\nu\rho} {\cal D}_{2}^{\mu}{\cal D}_{2}\cdot{\cal D}_{3}
+ \eta^{\nu\rho} {\cal D}_{3}^{\mu}{\cal D}_{2}\cdot{\cal D}_{3})
\nonumber \\
+ A_{9, abc} (\eta^{\mu\nu} {\cal D}_{1}^{\rho}{\cal D}_{1}\cdot{\cal D}_{3}
+ \eta^{\mu\nu} {\cal D}_{2}^{\rho}{\cal D}_{2}\cdot{\cal D}_{3}
+ \eta^{\mu\rho} {\cal D}_{1}^{\nu}{\cal D}_{1}\cdot{\cal D}_{2}
\nonumber \\
+ \eta^{\mu\rho} {\cal D}_{3}^{\nu}{\cal D}_{2}\cdot{\cal D}_{3}
+ \eta^{\nu\rho} {\cal D}_{2}^{\mu}{\cal D}_{1}\cdot{\cal D}_{2}
+ \eta^{\nu\rho} {\cal D}_{3}^{\mu}{\cal D}_{1}\cdot{\cal D}_{3})
\nonumber \\
+ A_{10, abc} (\eta^{\mu\nu} {\cal D}_{3}^{\rho}{\cal D}_{1}\cdot{\cal D}_{3}
+ \eta^{\mu\nu} {\cal D}_{3}^{\rho}{\cal D}_{2}\cdot{\cal D}_{3}
+ \eta^{\mu\rho} {\cal D}_{2}^{\nu}{\cal D}_{1}\cdot{\cal D}_{2}
\nonumber \\
+ \eta^{\mu\rho} {\cal D}_{2}^{\nu}{\cal D}_{1}\cdot{\cal D}_{3}
+ \eta^{\nu\rho} {\cal D}_{1}^{\mu}{\cal D}_{1}\cdot{\cal D}_{2}
+ \eta^{\nu\rho} {\cal D}_{1}^{\mu}{\cal D}_{1}\cdot{\cal D}_{3})
\nonumber \\
+ A_{11, abc} (\eta^{\mu\nu} {\cal D}_{2}^{\rho}{\cal D}_{1}\cdot{\cal D}_{3}
+ \eta^{\mu\nu} {\cal D}_{1}^{\rho}{\cal D}_{2}\cdot{\cal D}_{3}
+ \eta^{\mu\rho} {\cal D}_{3}^{\nu}{\cal D}_{1}\cdot{\cal D}_{2}
\nonumber \\
+ \eta^{\mu\rho} {\cal D}_{1}^{\nu}{\cal D}_{2}\cdot{\cal D}_{3}
+ \eta^{\nu\rho} {\cal D}_{3}^{\mu}{\cal D}_{1}\cdot{\cal D}_{2}
+ \eta^{\nu\rho} {\cal D}_{2}^{\mu}{\cal D}_{1}\cdot{\cal D}_{3})
\nonumber \\
+ 2 A_{12, abc} (\eta^{\mu\nu} {\cal D}_{3}^{\rho}{\cal D}_{1}\cdot{\cal D}_{2}
+ \eta^{\mu\rho} {\cal D}_{2}^{\nu}{\cal D}_{1}\cdot{\cal D}_{3}
+ \eta^{\nu\rho} {\cal D}_{1}^{\mu}{\cal D}_{2}\cdot{\cal D}_{3})]d(x,y,z)
\nonumber \\
u_{a}(x)~u_{b}(y)~u_{c}(z)
\eea
This expression is invariant (up to a sign) to the transformations
\be
x \leftrightarrow y, \mu \leftrightarrow \nu, 
{\cal D}_{1} \mapsto - {\cal D}_{2}, {\cal D}_{2} \mapsto - {\cal D}_{1},
{\cal D}_{3} \mapsto - {\cal D}_{3}
\ee
and
\be
y \leftrightarrow z, \nu \leftrightarrow \rho, 
{\cal D}_{1} \mapsto - {\cal D}_{1}, {\cal D}_{2} \mapsto - {\cal D}_{3},
{\cal D}_{3} \mapsto - {\cal D}_{2}
\ee
as we have explained in the preceding Section - see formulas (\ref{dABC1}) and (\ref{dABC2}). 

By direct computation we have the non-zero expressions
\bea
A_{2, abc} = f^{(0)}_{abc}, \quad A_{4, abc} = f^{(0)}_{abc} +  f^{(3)}_{abc},
\quad 
A_{5, abc} = {1 \over 3} (- f^{(0)}_{abc} +  f^{(4)}_{abc}),
\nonumber \\
A_{6, abc} = f^{(0)}_{abc} -  f^{(4)}_{abc},\quad
A_{7, abc} = {1 \over 3} (f^{(3)}_{abc} - 2 f^{(4)}_{abc}),
\quad
A_{8, abc} = f^{(0)}_{abc}, 
\nonumber \\
A_{10, abc} = - f^{(0)}_{abc},
\quad
A_{11, abc} = 2 f^{(4)}_{abc},\quad A_{12, abc} = - f^{(4)}_{abc},
\eea
where
\be
f^{(0)}_{[abc]} \equiv f_{eap} f_{ebq} f_{cpq}
\ee
\be
f^{(3)}_{[abc]} \equiv f^{\prime}_{epa} f^{\prime}_{eqb} f^{\prime}_{pqc}
\ee
\be
f^{(4)}_{[abc]} \equiv 
- i~Tr([t_{a}^{\epsilon},t_{b}^{\epsilon}]t_{c}^{\epsilon})
\ee
If we define
\be
g_{ab} = f_{apq}~f_{bpq}
\ee
\be
g^{(1)}_{ab} = f^{\prime}_{pqa}~f^{\prime}_{pqa}
\ee
\be
g^{(2)}_{ab} = \sum_{\epsilon} Tr(t^{\epsilon}_{a} t^{\epsilon}_{b})
\ee
then we have the alternative expressions:
\be
f^{(0)}_{[abc]} = {1\over 2} f_{abd}~g_{cd}
\ee
\be
f^{(3)}_{[abc]} = {1\over 2} f_{abd}~g^{(1)}_{cd}
\ee
\be
f^{(4)}_{[abc]} = f_{abd}~g^{(2)}_{cd}.
\ee
%\newpage
In the same way we derive consider the causal commutator
\bea
D^{[\mu\nu][\rho]\emptyset}(x,y,z) = D(T^{\mu\nu}(x),T^{\rho}(y);T(z)) 
\nonumber \\
- [ \bar{T}(T^{\mu\nu}(x),T^{\rho}(y)),T(z) ]
- [ T(T^{\mu\nu}(x),T(z)),T^{\rho}(y) ]
- [ T(T^{\rho}(y),T(z)),T^{\mu\nu}(x) ].
\eea
The triangle contribution has the form
\bea
D_{\rm triangle}^{[\mu\nu][\rho]\emptyset}(x,y,z) = i
( B_{1, abc} {\cal D}_{1}^{\mu}{\cal D}_{1}^{\rho}{\cal D}_{2}^{\nu}
+ B_{2, abc} {\cal D}_{1}^{\mu}{\cal D}_{1}^{\rho}{\cal D}_{3}^{\nu}
+ B_{3, abc} {\cal D}_{2}^{\mu}{\cal D}_{2}^{\rho}{\cal D}_{1}^{\nu}
\nonumber \\
+ B_{4, abc} {\cal D}_{2}^{\mu}{\cal D}_{2}^{\rho}{\cal D}_{3}^{\nu}
+ B_{5, abc} {\cal D}_{3}^{\mu}{\cal D}_{3}^{\rho}{\cal D}_{1}^{\nu}
+ B_{6, abc} {\cal D}_{3}^{\mu}{\cal D}_{3}^{\rho}{\cal D}_{2}^{\nu}
\nonumber \\
+ B_{7, abc} {\cal D}_{1}^{\mu}{\cal D}_{2}^{\nu}{\cal D}_{3}^{\rho}
+ B_{8, abc} {\cal D}_{1}^{\nu}{\cal D}_{2}^{\rho}{\cal D}_{3}^{\nu}
+ B_{9, abc} {\cal D}_{1}^{\rho}{\cal D}_{2}^{\mu}{\cal D}_{3}^{\nu})
\nonumber \\
+ B_{10, abc} \eta^{\mu\rho} {\cal D}_{1}^{\nu} {\cal D}_{1}\cdot{\cal D}_{2}
+ B_{11, abc} \eta^{\mu\rho} {\cal D}_{1}^{\nu} {\cal D}_{1}\cdot{\cal D}_{3}
+ B_{12, abc} \eta^{\mu\rho} {\cal D}_{2}^{\nu} {\cal D}_{1}\cdot{\cal D}_{2}
\nonumber \\
+ B_{13, abc} \eta^{\mu\rho} {\cal D}_{2}^{\nu} {\cal D}_{2}\cdot{\cal D}_{3}
+ B_{14, abc} \eta^{\mu\rho} {\cal D}_{3}^{\nu} {\cal D}_{1}\cdot{\cal D}_{3}
+ B_{15, abc} \eta^{\mu\rho} {\cal D}_{3}^{\nu} {\cal D}_{2}\cdot{\cal D}_{3}
\nonumber \\
+ B_{16, abc} \eta^{\mu\rho} {\cal D}_{1}^{\nu} {\cal D}_{2}\cdot{\cal D}_{3}
+ B_{17, abc} \eta^{\mu\rho} {\cal D}_{2}^{\nu} {\cal D}_{1}\cdot{\cal D}_{3}
+ B_{18, abc} \eta^{\mu\rho} {\cal D}_{3}^{\nu} {\cal D}_{1}\cdot{\cal D}_{2})
d(x,y,z)
\nonumber \\
u_{a}(x)~u_{b}(y)~u_{c}(z) - (\mu \leftrightarrow \nu)
\eea
where the non-zero 
$
B_{j}
$
are:
\be
B_{6, abc} = - f^{(0)}_{[abc]},\quad  B_{7, abc} = f^{(0)}_{[abc]}, \quad
B_{8, abc} = - f^{(0)}_{[abc]}, \quad  B_{9, abc} = f^{(0)}_{[abc]}\quad
B_{15, abc} = -f^{(0)}_{[abc]}. 
\ee
\newpage
\subsection{The Generic Form of the Coboundaries}
The generic expression 
$
B_{1}^{[\mu][\nu][\rho\sigma]}(x,y,z)
$
compatible with the restrictions is:
\be
B_{1}^{[\mu][\nu][\rho\sigma]}(x,y,z) = \sum_{j = 1}^{13}
a_{j,abc}~b_{j}^{[\mu][\nu][\rho\sigma]}(x,y,z) 
u_{a}(x) u_{b}(y) u_{c}(z)
\ee
where
\bea
b_{1}^{[\mu][\nu][\rho\sigma]}(x,y,z) = 
(\eta^{\mu\rho}{\cal D}_{1}^{\nu} {\cal D}_{1}^{\sigma}
- \eta^{\mu\sigma}{\cal D}_{1}^{\nu} {\cal D}_{1}^{\rho}
- \eta^{\nu\rho}{\cal D}_{2}^{\mu} {\cal D}_{2}^{\sigma}
+ \eta^{\nu\sigma}{\cal D}_{2}^{\mu} {\cal D}_{2}^{\rho})d(x,y,z)
\nonumber \\
b_{2}^{[\mu][\nu][\rho\sigma]}(x,y,z) = 
(\eta^{\nu\rho}{\cal D}_{1}^{\mu} {\cal D}_{1}^{\sigma}
- \eta^{\nu\sigma}{\cal D}_{1}^{\mu} {\cal D}_{1}^{\rho}
- \eta^{\mu\rho}{\cal D}_{2}^{\nu} {\cal D}_{2}^{\sigma}
+ \eta^{\mu\sigma}{\cal D}_{2}^{\nu} {\cal D}_{2}^{\rho})d(x,y,z)
\nonumber \\
b_{3}^{[\mu][\nu][\rho\sigma]}(x,y,z) = 
(\eta^{\mu\rho}{\cal D}_{1}^{\nu} {\cal D}_{2}^{\sigma}
- \eta^{\mu\sigma}{\cal D}_{1}^{\nu} {\cal D}_{2}^{\rho}
- \eta^{\nu\rho}{\cal D}_{1}^{\sigma} {\cal D}_{2}^{\mu}
+ \eta^{\nu\sigma}{\cal D}_{1}^{\rho} {\cal D}_{2}^{\mu})d(x,y,z)
\nonumber \\
b_{4}^{[\mu][\nu][\rho\sigma]}(x,y,z) = 
(\eta^{\mu\rho}{\cal D}_{1}^{\sigma} {\cal D}_{2}^{\nu}
- \eta^{\mu\sigma}{\cal D}_{1}^{\rho} {\cal D}_{2}^{\nu}
- \eta^{\nu\rho}{\cal D}_{1}^{\mu} {\cal D}_{2}^{\sigma}
+ \eta^{\nu\sigma}{\cal D}_{1}^{\mu} {\cal D}_{2}^{\rho})d(x,y,z)
\nonumber \\
b_{5}^{[\mu][\nu][\rho\sigma]}(x,y,z) = 
\eta^{\mu\nu}( {\cal D}_{1}^{\rho} {\cal D}_{2}^{\sigma}
- {\cal D}_{1}^{\sigma} {\cal D}_{2}^{\rho})d(x,y,z)
\nonumber \\
b_{6}^{[\mu][\nu][\rho\sigma]}(x,y,z) = 
( \eta^{\mu\rho}~\eta^{\nu\sigma} - \eta^{\mu\sigma}~\eta^{\nu\rho})
{\cal D}_{1} \cdot {\cal D}_{2}d(x,y,z)
\nonumber \\
b_{7}^{[\mu][\nu][\rho\sigma]}(x,y,z) = 
(\eta^{\mu\rho}{\cal D}_{1}^{\nu} {\cal D}_{3}^{\sigma}
- \eta^{\mu\sigma}{\cal D}_{1}^{\nu} {\cal D}_{3}^{\rho}
- \eta^{\nu\rho}{\cal D}_{2}^{\mu} {\cal D}_{3}^{\sigma}
+ \eta^{\nu\sigma}{\cal D}_{2}^{\mu} {\cal D}_{3}^{\rho})d(x,y,z)
\nonumber \\
b_{8}^{[\mu][\nu][\rho\sigma]}(x,y,z) = 
(\eta^{\mu\rho}{\cal D}_{1}^{\sigma} {\cal D}_{3}^{\nu}
- \eta^{\mu\sigma}{\cal D}_{1}^{\rho} {\cal D}_{3}^{\nu}
- \eta^{\nu\rho}{\cal D}_{2}^{\sigma} {\cal D}_{3}^{\mu}
+ \eta^{\nu\sigma}{\cal D}_{2}^{\rho} {\cal D}_{3}^{\mu})d(x,y,z)
\nonumber \\
b_{9}^{[\mu][\nu][\rho\sigma]}(x,y,z) = 
(\eta^{\nu\rho}{\cal D}_{1}^{\mu} {\cal D}_{3}^{\sigma}
- \eta^{\nu\sigma}{\cal D}_{1}^{\mu} {\cal D}_{3}^{\rho}
- \eta^{\mu\rho}{\cal D}_{2}^{\nu} {\cal D}_{3}^{\sigma}
+ \eta^{\mu\sigma}{\cal D}_{2}^{\nu} {\cal D}_{3}^{\rho})d(x,y,z)
\nonumber \\
b_{10}^{[\mu][\nu][\rho\sigma]}(x,y,z) = 
(\eta^{\nu\rho}{\cal D}_{1}^{\sigma} {\cal D}_{3}^{\mu}
- \eta^{\nu\sigma}{\cal D}_{1}^{\rho} {\cal D}_{3}^{\mu}
- \eta^{\mu\rho}{\cal D}_{2}^{\sigma} {\cal D}_{3}^{\nu}
+ \eta^{\mu\sigma}{\cal D}_{2}^{\rho} {\cal D}_{3}^{\nu})d(x,y,z)
\nonumber \\
b_{11}^{[\mu][\nu][\rho\sigma]}(x,y,z) = 
\eta^{\mu\nu}( {\cal D}_{1}^{\rho} {\cal D}_{3}^{\sigma}
- {\cal D}_{1}^{\sigma} {\cal D}_{3}^{\rho}
- {\cal D}_{2}^{\rho} {\cal D}_{3}^{\sigma}
+ {\cal D}_{2}^{\sigma} {\cal D}_{3}^{\rho})d(x,y,z)
\nonumber \\
b_{12}^{[\mu][\nu][\rho\sigma]}(x,y,z) = 
( \eta^{\mu\rho}~\eta^{\nu\sigma} - \eta^{\mu\sigma}~\eta^{\nu\rho})
( {\cal D}_{1} \cdot {\cal D}_{3} + {\cal D}_{2} \cdot {\cal D}_{3})d(x,y,z)
\nonumber \\
b_{13}^{[\mu][\nu][\rho\sigma]}(x,y,z) = 
(\eta^{\mu\rho}{\cal D}_{3}^{\nu} {\cal D}_{3}^{\sigma}
- \eta^{\mu\sigma}{\cal D}_{3}^{\nu} {\cal D}_{3}^{\rho}
- \eta^{\nu\rho}{\cal D}_{3}^{\mu} {\cal D}_{3}^{\sigma}
+ \eta^{\nu\sigma}{\cal D}_{3}^{\mu} {\cal D}_{3}^{\rho})d(x,y,z)
\eea
\newpage
The equation
\be
D_{\rm triangle}^{[\mu][\nu][\rho]}(x,y,z)_{0} = (\bar{s}B)^{[\mu][\nu][\rho]}(x,y,z) 
+ D_{\delta}^{[\mu][\nu][\rho]}(x,y,z)
\ee
gives the following equations in the sector
$
u_{a}(x)~u_{b}(y)~u_{c}(z)
$:
\bea
a_{1} + a_{2} = A_{1}
\nonumber \\
a_{3} - a_{4} + a_{8} - a_{13} = A_{2}
\nonumber \\
- a_{2} + a_{4} + a_{7} + a_{9} + a_{13} = A_{3}
\nonumber \\
- a_{1} - a_{3} + a_{10} = A_{4}
\nonumber \\
- a_{9} = A_{5}
\nonumber \\
- a_{7} - a_{10} = A_{6}
\nonumber \\
- a_{8} = A_{7}
\nonumber \\
- a_{5} - a_{7} - a_{12} - a_{13} = A_{8}
\nonumber \\
a_{1} - a_{3} - a_{6} - a_{11} + a_{13} = A_{9}
\nonumber \\
a_{2} + a_{4} + a_{6} - a_{9} + a_{12} = A_{10}
\nonumber \\
a_{7} + a_{8} + a_{10} + a_{11} - a_{12} = A_{11}
\nonumber \\
a_{9} - a_{11} + a_{12} = A_{12}
\label{a}
\eea
If we solve the system we obtain the following consistency equations
\bea
A_{5} + A_{6} + A_{7} + A_{11} + A_{12} = 0
\nonumber \\
A_{1} + A_{2} + A_{3} + A_{4} + A_{5} + A_{6} + A_{7} = 0
\nonumber \\
A_{1} - A_{2} + 2 A_{5} - A_{7} - A_{9} - A_{10} + A_{12} = 0
\eea
If we use the explicit expressions for
$
A_{j},~j = 1,\dots,12
$
we obtain the equation
\be
2 f^{(0)}_{[abc]} + f^{(3)}_{[abc]} - f^{(4)}_{[abc]} = 0
\label{fff1}
\ee
and this is one of the restrictions on the constants from the Lagrangian which are
necessary to have the super-renormalizability property.
\newpage
In the same way we have the generic form
\be
B_{1}^{[\mu\nu][\rho\sigma]\emptyset}(x,y,z) = \sum_{j = 1}^{13}
b_{j,abc}~b_{j}^{[\mu\nu][\rho\sigma]\emptyset}(x,y,z)
u_{a}(x) u_{b}(y) u_{c}(z)
\ee
where
\bea
b_{1}^{[\mu\nu][\rho\sigma]\emptyset}(x,y,z) = 
(\eta^{\mu\rho}{\cal D}_{1}^{\nu} {\cal D}_{1}^{\sigma}
- \eta^{\nu\rho}{\cal D}_{1}^{\mu} {\cal D}_{1}^{\sigma}
- \eta^{\mu\sigma}{\cal D}_{1}^{\nu} {\cal D}_{1}^{\rho}
+ \eta^{\nu\sigma}{\cal D}_{1}^{\mu} {\cal D}_{1}^{\rho}
\nonumber \\
+ \eta^{\mu\rho}{\cal D}_{2}^{\nu} {\cal D}_{2}^{\sigma}
- \eta^{\mu\sigma}{\cal D}_{2}^{\nu} {\cal D}_{2}^{\rho}
- \eta^{\nu\rho}{\cal D}_{2}^{\mu} {\cal D}_{2}^{\sigma}
+ \eta^{\nu\sigma}{\cal D}_{2}^{\mu} {\cal D}_{2}^{\rho})d(x,y,z)
\nonumber \\
b_{2}^{[\mu\nu][\rho\sigma]\emptyset}(x,y,z) = 
(\eta^{\mu\rho}{\cal D}_{1}^{\nu} {\cal D}_{2}^{\sigma}
- \eta^{\nu\rho}{\cal D}_{1}^{\mu} {\cal D}_{2}^{\sigma}
- \eta^{\mu\sigma}{\cal D}_{1}^{\nu} {\cal D}_{2}^{\rho}
+ \eta^{\nu\sigma}{\cal D}_{1}^{\mu} {\cal D}_{2}^{\rho})d(x,y,z)
\nonumber \\
b_{3}^{[\mu\nu][\rho\sigma]\emptyset}(x,y,z) = 
(\eta^{\mu\rho}{\cal D}_{1}^{\sigma} {\cal D}_{2}^{\nu}
- \eta^{\nu\rho}{\cal D}_{1}^{\sigma} {\cal D}_{2}^{\mu}
- \eta^{\mu\sigma}{\cal D}_{1}^{\rho} {\cal D}_{2}^{\nu}
+ \eta^{\nu\sigma}{\cal D}_{1}^{\rho} {\cal D}_{2}^{\mu})d(x,y,z)
\nonumber \\
b_{4}^{[\mu\nu][\rho\sigma]\emptyset}(x,y,z) = 
( \eta^{\mu\rho}~\eta^{\nu\sigma} - \eta^{\mu\sigma}~\eta^{\nu\rho})
{\cal D}_{1} \cdot {\cal D}_{2}d(x,y,z)
\nonumber \\
b_{5}^{[\mu\nu][\rho\sigma]\emptyset}(x,y,z) = 
(\eta^{\mu\rho}{\cal D}_{1}^{\nu} {\cal D}_{3}^{\sigma}
- \eta^{\nu\rho}{\cal D}_{1}^{\mu} {\cal D}_{3}^{\sigma}
- \eta^{\mu\sigma}{\cal D}_{1}^{\nu} {\cal D}_{3}^{\rho}
+ \eta^{\nu\sigma}{\cal D}_{1}^{\mu} {\cal D}_{3}^{\rho}
\nonumber \\
+ \eta^{\mu\rho}{\cal D}_{2}^{\sigma} {\cal D}_{3}^{\nu}
- \eta^{\mu\sigma}{\cal D}_{2}^{\rho} {\cal D}_{3}^{\nu}
- \eta^{\nu\rho}{\cal D}_{2}^{\sigma} {\cal D}_{3}^{\mu}
+ \eta^{\nu\sigma}{\cal D}_{2}^{\rho} {\cal D}_{3}^{\mu})d(x,y,z)
\nonumber \\
b_{6}^{[\mu\nu][\rho\sigma]\emptyset}(x,y,z) = 
(\eta^{\mu\rho}{\cal D}_{1}^{\sigma} {\cal D}_{3}^{\nu}
- \eta^{\nu\rho}{\cal D}_{1}^{\sigma} {\cal D}_{3}^{\mu}
- \eta^{\mu\sigma}{\cal D}_{1}^{\rho} {\cal D}_{3}^{\nu}
+ \eta^{\nu\sigma}{\cal D}_{1}^{\rho} {\cal D}_{3}^{\mu}
\nonumber \\
+ \eta^{\mu\rho}{\cal D}_{2}^{\nu} {\cal D}_{3}^{\sigma}
- \eta^{\mu\sigma}{\cal D}_{2}^{\nu} {\cal D}_{3}^{\rho}
- \eta^{\nu\rho}{\cal D}_{2}^{\mu} {\cal D}_{3}^{\sigma}
+ \eta^{\nu\sigma}{\cal D}_{2}^{\mu} {\cal D}_{3}^{\rho})d(x,y,z)
\nonumber \\
b_{7}^{[\mu\nu][\rho\sigma]\emptyset}(x,y,z) =
( \eta^{\mu\rho}~\eta^{\nu\sigma} - \eta^{\mu\sigma}~\eta^{\nu\rho})
( {\cal D}_{1} \cdot {\cal D}_{3} + {\cal D}_{2} \cdot {\cal D}_{3})d(x,y,z)
\nonumber \\
b_{8}^{[\mu\nu][\rho\sigma]\emptyset}(x,y,z) = 
(\eta^{\mu\rho}{\cal D}_{3}^{\nu} {\cal D}_{3}^{\sigma}
- \eta^{\nu\rho}{\cal D}_{3}^{\mu} {\cal D}_{3}^{\sigma}
- \eta^{\mu\sigma}{\cal D}_{3}^{\nu} {\cal D}_{3}^{\rho}
+ \eta^{\nu\sigma}{\cal D}_{3}^{\mu} {\cal D}_{3}^{\rho})d(x,y,z)
\eea
%\newpage
and the equation
\be
D_{\rm triangle}^{[\mu\nu][\rho]\emptyset}(x,y,z)_{0} = 
(\bar{s}B)^{[\mu\nu][\rho]\emptyset}(x,y,z) 
+ D_{\delta}^{[\mu\nu][\rho]\emptyset}(x,y,z)
\ee
gives the following equations in the sector
$
u_{a}(x)~u_{b}(y)~u_{c}(z)
$:
\bea
a_{10} - a_{11} - a_{13} - b_{3} = B_{1}
\nonumber \\
- a_{8} + a_{11} - b_{1} - b_{6} = B_{2}
\nonumber \\
- a_{2} - a_{9} - a_{11} + b_{1} = B_{3}
\nonumber \\
a_{4} - a_{5} - b_{1} = B_{4}
\nonumber \\
a_{1} + b_{5} + b_{8} = B_{5}
\nonumber \\
- a_{1} + b_{6} = B_{6}
\nonumber \\
- a_{3} - a_{7} - b_{6} = B_{7}
\nonumber \\
- a_{4} - a_{11} - b_{2} - b_{5} = B_{8}
\nonumber \\
a_{5} + a_{8} - b_{3} = B_{9}
\nonumber \\
a_{7} + a_{12} - a_{13} + b_{2} + b_{4} = B_{10}
\nonumber \\
- a_{9} + a_{12} - b_{1} + b_{5} + b_{7} = B_{11}
\nonumber \\
a_{1} - a_{8} - a_{12} + b_{1} = B_{12}
\nonumber \\
- a_{4} - a_{6} - b_{1} = B_{13}
\nonumber \\
- a_{2} - b_{6} - b_{7} + b_{8} = B_{14}
\nonumber \\
a_{2} - b_{5} - b_{7} = B_{15}
\nonumber \\
a_{6} + a_{9} - b_{2} + b_{7} = B_{16}
\nonumber \\
a_{4} - a_{12} - b_{3} + b_{6} = B_{17}
\nonumber \\
a_{3} + a_{10} - b_{4} + b_{5} = B_{18}
\label{a+b}
\eea

The systems (\ref{a}) and (\ref{a+b}) can be used to obtain the parameters $a$
and $b$ but they do not produce new restrictions on $A$ and $B$. So if we impose 
the restriction (\ref{fff1}) we have the super-renormalizability property in the 
top ghost number. We can use a descent procedure to show that this property
stays true for all triangle contributions of
$
D^{IJK}
$
without derivatives on the fields. So, if we consider the expression
\be
D^{IJK} - (\bar{s}B_{1})^{IJK}
\label{D2}
\ee
we eliminate all the terms without derivatives on the fields.
\newpage
\subsection{The Causal Commutators for Ghost Number $2$}

There are two relevant causal commutators of this type:
\bea
D^{\emptyset\emptyset[\mu\nu]}(x,y,z) = D(T(x),T(y),T^{\mu\nu}(z)) 
\nonumber \\
- [ \bar{T}(x,y),T^{\mu\nu}(y)),T(z) ]
- [ T^{\emptyset[\mu\nu]}(x,z),T(z)),T(y) ]
- [ T^{\emptyset[\mu\nu]}(y,z),T(x) ]
\eea
and 
\bea
D^{[\mu][\nu]\emptyset}(x,y,z) = D(T^{\mu}(x),T^{\nu}(y),T(z)) 
\nonumber \\
- [ \bar{T}^{[\mu][\nu]}(x,y)),T(z) ]
- [ T^{[\mu]\emptyset}(x,z),T^{\nu}(y) ]
+ [ T^{[\nu]\emptyset}(y,z),T^{\mu}(x) ].
\eea
Both expressions have a contribution
$
D_{1} \sim uu v
$
and a contribution
$
D_{2} \sim u u F
$.
The contributions of the second type are by explicit computation:
\bea
D_{2}^{\emptyset\emptyset[\mu\nu]}(x,y,z) = i~f_{abc}^{(0)}
\nonumber \\
~[ {\cal D}_{1\rho} {\cal D}_{2}^{\nu}d(x,y,z) W_{1}^{\mu\rho}(x,y,z)
- {\cal D}_{1}^{\nu} {\cal D}_{2\rho}d(x,y,z) W_{2}^{\mu\rho}(x,y,z) ]
- (\mu \leftrightarrow \nu)
\nonumber \\
+ i {\cal D}_{1} \cdot {\cal D}_{2}d(x,y,z) W_{3}^{\mu\rho}(x,y,z)
\eea
and 
\bea
D_{2}^{\emptyset\emptyset[\mu\nu]}(x,y,z) = i~f_{abc}^{(0)}
\nonumber \\
~[ - {\cal D}_{2} \cdot {\cal D}_{3}d(x,y,z) W_{1}^{\mu\nu}(x,y,z)
+ {\cal D}_{1} \cdot {\cal D}_{3}d(x,y,z) W_{2}^{\mu\nu}(x,y,z)
\nonumber \\
+ ( - {\cal D}_{1\rho} {\cal D}_{2}^{\nu} + {\cal D}_{1}^{\nu} {\cal D}_{2\rho}
+ {\cal D}_{2\rho} {\cal D}_{3}^{\nu} - {\cal D}_{1\rho} {\cal D}_{3}^{\nu})
d(x,y,z) W_{1}^{\mu\rho}(x,y,z)
\nonumber \\
+ ( - {\cal D}_{1}^{\mu} {\cal D}_{2\rho} + {\cal D}_{1\rho} {\cal D}_{2}^{\mu}
+ {\cal D}_{2\rho} {\cal D}_{3}^{\mu} - {\cal D}_{2\rho} {\cal D}_{3}^{\mu})
d(x,y,z) W_{2}^{\nu\rho}(x,y,z)
\nonumber \\
- {\cal D}_{1} \cdot {\cal D}_{2}d(x,y,z) W_{3}^{\mu\nu}(x,y,z)
\nonumber \\
- {\cal D}_{1\rho} {\cal D}_{2}^{\mu}d(x,y,z) W_{3}^{\nu\rho}(x,y,z)
+ {\cal D}_{1}^{\nu} {\cal D}_{2\rho}d(x,y,z) W_{3}^{\mu\rho}(x,y,z)
\nonumber \\
+ ({\cal D}_{2}^{\mu} {\cal D}_{3\rho} - {\cal D}_{1\rho} {\cal D}_{3}^{\mu}
+ {\cal D}_{3}^{\mu} {\cal D}_{3\rho} + {\cal D}_{1}^{\mu} {\cal D}_{3\rho})
d(x,y,z) W_{3}^{\nu\rho}(x,y,z)
\nonumber \\
+ ( - {\cal D}_{1}^{\nu} {\cal D}_{3\rho} + {\cal D}_{2\rho} {\cal D}_{3}^{\nu}
- {\cal D}_{3}^{\nu} {\cal D}_{3\rho} - {\cal D}_{2}^{\nu} {\cal D}_{3\rho})
d(x,y,z) W_{3}^{\mu\rho}(x,y,z)]
\eea
where
\be
W_{1}^{\mu\nu} \equiv F_{a}^{\mu\nu}(x) u_{b}(y) u_{c}(z), \quad
W_{2}^{\mu\nu} \equiv u_{a}(x) F_{b}^{\mu\nu}(y) u_{c}(z), \quad
W_{3}^{\mu\nu} \equiv u_{a}(x) u_{b}(y) F_{c}^{\mu\nu}(z)
\ee
In the expressions (\ref{D2}) we have some supplementary terms of the type
$
u u F
$
coming from
$
- (\bar{s}B_{1})^{IJK}
$.
The coboundary equations are in this case:
\be
D_{2}^{\emptyset\emptyset[\mu\nu]} 
+ {i \over 2} \sum [ b_{j,abc} b_{j}^{[\rho\sigma]\emptyset [\mu\nu]}
W_{1,\rho\sigma} + (x \leftrightarrow y) ] = 
(\bar{s}B_{2})^{\emptyset\emptyset[\mu\nu]}
+ D_{\delta}^{\emptyset\emptyset[\mu\nu]}
\label{co1}
\ee 
and 
\be
D^{[\mu][\nu]\emptyset} 
+ {i \over 2} \sum_{j} a_{j,abc} b_{j}^{[\mu][\nu][\rho\sigma]}
W_{3,\rho\sigma} = 
(\bar{s}B_{2})^{[\mu][\nu]\emptyset}
+ D_{\delta}^{[\mu][\nu]\emptyset}.
\label{co2}
\ee 
\newpage
\subsection{The Generic Form of the Coboundaries}
The various restrictions lead to the following generic forms:
\be
B_{2}^{[\rho]\emptyset[\mu\nu]}(x,y,z) = \sum_{j=1}^{30}
c_{j,abc}~B_{j}^{[\rho]\emptyset[\mu\nu]}(x,y,z)
\ee
where
\bea
B_{1}^{[\rho]\emptyset[\mu\nu]}(x,y,z) = {\cal D}_{1}^{\rho}d(x,y,z)
W_{1}^{\mu\nu}
\nonumber \\
B_{2}^{[\rho]\emptyset[\mu\nu]}(x,y,z) = {\cal D}_{1}^{\rho}d(x,y,z)
W_{2}^{\mu\nu}
\nonumber \\
B_{1}^{[\rho]\emptyset[\mu\nu]}(x,y,z) = {\cal D}_{1}^{\rho}d(x,y,z)
W_{3}^{\mu\nu}
\nonumber \\
B_{4}^{[\rho]\emptyset[\mu\nu]}(x,y,z) = {\cal D}_{2}^{\rho}d(x,y,z)
W_{1}^{\mu\nu}
\nonumber \\
B_{5}^{[\rho]\emptyset[\mu\nu]}(x,y,z) = {\cal D}_{2}^{\rho}d(x,y,z)
W_{2}^{\mu\nu}
\nonumber \\
B_{6}^{[\rho]\emptyset[\mu\nu]}(x,y,z) = {\cal D}_{2}^{\rho}d(x,y,z)
W_{3}^{\mu\nu}
\nonumber \\
B_{7}^{[\rho]\emptyset[\mu\nu]}(x,y,z) = {\cal D}_{3}^{\rho}d(x,y,z)
W_{1}^{\mu\nu}
\nonumber \\
B_{8}^{[\rho]\emptyset[\mu\nu]}(x,y,z) = {\cal D}_{3}^{\rho}d(x,y,z)
W_{2}^{\mu\nu}
\nonumber \\
B_{9}^{[\rho]\emptyset[\mu\nu]}(x,y,z) = {\cal D}_{3}^{\rho}d(x,y,z)
W_{3}^{\mu\nu}
\nonumber \\
B_{10}^{[\rho]\emptyset[\mu\nu]}(x,y,z) = {\cal D}_{1}^{\nu}d(x,y,z)
W_{1}^{\mu\rho} - (\mu \leftrightarrow \nu)
\nonumber \\
B_{11}^{[\rho]\emptyset[\mu\nu]}(x,y,z) = {\cal D}_{1}^{\nu}d(x,y,z)
W_{2}^{\mu\rho} - (\mu \leftrightarrow \nu)
\nonumber \\
B_{12}^{[\rho]\emptyset[\mu\nu]}(x,y,z) = {\cal D}_{1}^{\nu}d(x,y,z)
W_{3}^{\mu\rho} - (\mu \leftrightarrow \nu)
\nonumber \\
B_{13}^{[\rho]\emptyset[\mu\nu]}(x,y,z) = {\cal D}_{2}^{\nu}d(x,y,z)
W_{1}^{\mu\rho} - (\mu \leftrightarrow \nu)
\nonumber \\
B_{14}^{[\rho]\emptyset[\mu\nu]}(x,y,z) = {\cal D}_{2}^{\nu}d(x,y,z)
W_{2}^{\mu\rho} - (\mu \leftrightarrow \nu)
\nonumber \\
B_{15}^{[\rho]\emptyset[\mu\nu]}(x,y,z) = {\cal D}_{2}^{\nu}d(x,y,z)
W_{3}^{\mu\rho} - (\mu \leftrightarrow \nu)
\nonumber \\
B_{16}^{[\rho]\emptyset[\mu\nu]}(x,y,z) = {\cal D}_{3}^{\nu}d(x,y,z)
W_{1}^{\mu\rho} - (\mu \leftrightarrow \nu)
\nonumber \\
B_{17}^{[\rho]\emptyset[\mu\nu]}(x,y,z) = {\cal D}_{3}^{\nu}d(x,y,z)
W_{2}^{\mu\rho} - (\mu \leftrightarrow \nu) 
\nonumber \\
B_{18}^{[\rho]\emptyset[\mu\nu]}(x,y,z) = {\cal D}_{3}^{\nu}d(x,y,z)
W_{3}^{\mu\rho}- (\mu \leftrightarrow \nu)
\nonumber \\
B_{19}^{[\rho]\emptyset[\mu\nu]}(x,y,z) = \eta^{\nu\rho}
{\cal D}_{1\sigma}d(x,y,z)
W_{1}^{\mu\sigma} - (\mu \leftrightarrow \nu)
\nonumber \\
B_{20}^{[\rho]\emptyset[\mu\nu]}(x,y,z) = \eta^{\nu\rho}
{\cal D}_{1\sigma}d(x,y,z)
W_{2}^{\mu\sigma} - (\mu \leftrightarrow \nu)
\nonumber \\
B_{21}^{[\rho]\emptyset[\mu\nu]}(x,y,z) = \eta^{\nu\rho}
{\cal D}_{1\sigma}d(x,y,z)
W_{3}^{\mu\sigma} - (\mu \leftrightarrow \nu)
\nonumber \\
B_{22}^{[\rho]\emptyset[\mu\nu]}(x,y,z) = \eta^{\nu\rho}
{\cal D}_{2\sigma}d(x,y,z)
W_{1}^{\mu\sigma} - (\mu \leftrightarrow \nu)
\nonumber \\
B_{23}^{[\rho]\emptyset[\mu\nu]}(x,y,z) = \eta^{\nu\rho}
{\cal D}_{2\sigma}d(x,y,z)
W_{2}^{\mu\sigma} - (\mu \leftrightarrow \nu)
\nonumber \\
B_{24}^{[\rho]\emptyset[\mu\nu]}(x,y,z) = \eta^{\nu\rho}
{\cal D}_{2\sigma}d(x,y,z)
W_{3}^{\mu\sigma} - (\mu \leftrightarrow \nu)
\nonumber \\
B_{25}^{[\rho]\emptyset[\mu\nu]}(x,y,z) = \eta^{\nu\rho}
{\cal D}_{3\sigma}d(x,y,z)
W_{1}^{\mu\sigma} - (\mu \leftrightarrow \nu)
\nonumber \\
B_{26}^{[\rho]\emptyset[\mu\nu]}(x,y,z) = \eta^{\nu\rho}
{\cal D}_{3\sigma}d(x,y,z)
W_{2}^{\mu\sigma} - (\mu \leftrightarrow \nu) 
\nonumber \\
B_{27}^{[\rho]\emptyset[\mu\nu]}(x,y,z) = \eta^{\nu\rho}
{\cal D}_{3\sigma}d(x,y,z)
W_{3}^{\mu\sigma}- (\mu \leftrightarrow \nu)
\eea
\bea
B_{28}^{[\rho]\emptyset[\mu\nu]}(x,y,z) = d(x,y,z)
\partial^{\rho}F_{a}^{\mu\nu}(x) u_{b}(y) u_{c}(z)
\nonumber \\
B_{29}^{[\rho]\emptyset[\mu\nu]}(x,y,z) = d(x,y,z)
u_{a}(x) \partial^{\rho}F_{b}^{\mu\nu}(y) u_{c}(z)
\nonumber \\
B_{30}^{[\rho]\emptyset[\mu\nu]}(x,y,z) = d(x,y,z)
u_{a}(x) u_{b}(y) \partial^{\rho}F_{c}^{\mu\nu}(z)
\eea
and
\be
B_{2}^{[\mu][\nu][\rho]}(x,y,z) = \sum_{j = 1}^{10}
d_{j,abc}~B_{j}^{[\mu][\nu][\rho]}(x,y,z)
\ee
where
\bea
B_{1}^{[\mu][\nu][\rho]}(x,y,z) = 
{\cal D}_{1}^{\rho}d(x,y,z) W_{1}^{\mu\nu}
+ {\cal D}_{2}^{\rho}d(x,y,z) W_{2}^{\mu\nu}
+ {\cal D}_{1}^{\nu}d(x,y,z) W_{1}^{\mu\rho}
\nonumber \\
- {\cal D}_{3}^{\nu}d(x,y,z) W_{3}^{\mu\rho}
- {\cal D}_{2}^{\mu}d(x,y,z) W_{2}^{\nu\rho}
- {\cal D}_{3}^{\mu}d(x,y,z) W_{3}^{\nu\rho}
\nonumber \\
B_{2}^{[\mu][\nu][\rho]}(x,y,z) = 
{\cal D}_{1}^{\rho}d(x,y,z) W_{2}^{\mu\nu}
+ {\cal D}_{2}^{\rho}d(x,y,z) W_{1}^{\mu\nu}
- {\cal D}_{1}^{\nu}d(x,y,z) W_{3}^{\mu\rho}
\nonumber \\
+ {\cal D}_{3}^{\nu}d(x,y,z) W_{1}^{\mu\rho}
- {\cal D}_{2}^{\mu}d(x,y,z) W_{3}^{\nu\rho}
- {\cal D}_{3}^{\mu}d(x,y,z) W_{2}^{\nu\rho}
\nonumber \\
B_{3}^{[\mu][\nu][\rho]}(x,y,z) = 
{\cal D}_{1}^{\rho}d(x,y,z) W_{3}^{\mu\nu}
- {\cal D}_{2}^{\rho}d(x,y,z) W_{3}^{\mu\nu}
- {\cal D}_{1}^{\nu}d(x,y,z) W_{2}^{\mu\rho}
\nonumber \\
+ {\cal D}_{3}^{\nu}d(x,y,z) W_{2}^{\mu\rho}
+ {\cal D}_{2}^{\mu}d(x,y,z) W_{1}^{\nu\rho}
- {\cal D}_{3}^{\mu}d(x,y,z) W_{1}^{\nu\rho}
\nonumber \\
B_{4}^{[\mu][\nu][\rho]}(x,y,z) = 
{\cal D}_{3}^{\rho}d(x,y,z) W_{1}^{\mu\nu}
+ {\cal D}_{3}^{\rho}d(x,y,z) W_{2}^{\mu\nu}
+ {\cal D}_{2}^{\nu}d(x,y,z) W_{1}^{\mu\rho}
\nonumber \\
- {\cal D}_{2}^{\nu}d(x,y,z) W_{3}^{\mu\rho}
- {\cal D}_{1}^{\mu}d(x,y,z) W_{2}^{\nu\rho}
- {\cal D}_{1}^{\mu}d(x,y,z) W_{3}^{\nu\rho}
\nonumber \\
B_{5}^{[\mu][\nu][\rho]}(x,y,z) = 
\eta^{\mu\rho} {\cal D}_{1\sigma}d(x,y,z) W_{1}^{\nu\sigma}
- \eta^{\nu\rho} {\cal D}_{2\sigma}d(x,y,z) W_{2}^{\mu\sigma}
+ \eta^{\mu\nu} {\cal D}_{1\sigma}d(x,y,z) W_{1}^{\rho\sigma}
\nonumber \\
+ \eta^{\nu\rho} {\cal D}_{3\sigma}d(x,y,z) W_{3}^{\mu\sigma}
- \eta^{\mu\nu} {\cal D}_{2\sigma}d(x,y,z) W_{2}^{\rho\sigma}
+ \eta^{\mu\rho} {\cal D}_{3\sigma}d(x,y,z) W_{3}^{\nu\sigma}
\nonumber \\
B_{6}^{[\mu][\nu][\rho]}(x,y,z) = 
\eta^{\nu\rho} {\cal D}_{1\sigma}d(x,y,z) W_{1}^{\mu\sigma}
- \eta^{\mu\rho} {\cal D}_{2\sigma}d(x,y,z) W_{2}^{\nu\sigma}
+ \eta^{\mu\nu} {\cal D}_{3\sigma}d(x,y,z) W_{3}^{\rho\sigma}
\nonumber \\
B_{7}^{[\mu][\nu][\rho]}(x,y,z) = 
\eta^{\mu\rho} {\cal D}_{1\sigma}d(x,y,z) W_{2}^{\nu\sigma}
- \eta^{\nu\rho} {\cal D}_{2\sigma}d(x,y,z) W_{1}^{\mu\sigma}
- \eta^{\mu\nu} {\cal D}_{1\sigma}d(x,y,z) W_{3}^{\rho\sigma}
\nonumber \\
- \eta^{\nu\rho} {\cal D}_{3\sigma}d(x,y,z) W_{1}^{\mu\sigma}
- \eta^{\mu\nu} {\cal D}_{2\sigma}d(x,y,z) W_{3}^{\rho\sigma}
+ \eta^{\mu\rho} {\cal D}_{3\sigma}d(x,y,z) W_{2}^{\nu\sigma}
\nonumber \\
B_{8}^{[\mu][\nu][\rho]}(x,y,z) = 
\eta^{\nu\rho} {\cal D}_{1\sigma}d(x,y,z) W_{2}^{\mu\sigma}
- \eta^{\mu\rho} {\cal D}_{2\sigma}d(x,y,z) W_{1}^{\nu\sigma}
- \eta^{\nu\rho} {\cal D}_{1\sigma}d(x,y,z) W_{3}^{\mu\sigma}
\nonumber \\
- \eta^{\mu\nu} {\cal D}_{3\sigma}d(x,y,z) W_{1}^{\mu\sigma}
- \eta^{\mu\rho} {\cal D}_{2\sigma}d(x,y,z) W_{3}^{\nu\sigma}
+ \eta^{\mu\nu} {\cal D}_{3\sigma}d(x,y,z) W_{2}^{\rho\sigma}
\nonumber \\
B_{9}^{[\mu][\nu][\rho]}(x,y,z) = 
\eta^{\mu\rho} {\cal D}_{1\sigma}d(x,y,z) W_{3}^{\nu\sigma}
+ \eta^{\nu\rho} {\cal D}_{2\sigma}d(x,y,z) W_{3}^{\mu\sigma}
- \eta^{\mu\nu} {\cal D}_{1\sigma}d(x,y,z) W_{2}^{\rho\sigma}
\nonumber \\
- \eta^{\nu\rho} {\cal D}_{3\sigma}d(x,y,z) W_{2}^{\mu\sigma}
+ \eta^{\mu\nu} {\cal D}_{2\sigma}d(x,y,z) W_{1}^{\rho\sigma}
+ \eta^{\mu\rho} {\cal D}_{3\sigma}d(x,y,z) W_{1}^{\nu\sigma}
\eea
\be
B_{10}^{[\mu][\nu][\rho]}(x,y,z) = d(x,y,z)
[ \partial^{\mu}F_{a}^{\nu\rho}(x) u_{b}(y) u_{c}(z)
+ u_{a}(x) \partial^{\nu}F_{b}^{\mu\rho}(y) u_{c}(z)
+ u_{a}(x) u_{b}(y) \partial^{\rho}F_{c}^{\mu\nu}(z) ]
\ee
\newpage

From the equation (\ref{co1})
\bea
D_{2}^{\emptyset\emptyset[\mu\nu]} 
+ {i \over 2} \sum [ b_{j,abc} b_{j}^{[\rho\sigma]\emptyset [\mu\nu]}
W_{1,\rho\sigma} + (x \leftrightarrow y) ] = 
(\bar{s}B_{2})^{\emptyset\emptyset[\mu\nu]}
+ D_{\delta}^{\emptyset\emptyset[\mu\nu]}
\nonumber
\eea 
we get in the sector
$
{\cal D}{\cal D}d(x,y,z) u u \partial F
$
\bea
c_{1} + c_{19} - c_{29} = 0
\nonumber \\
c_{4} + c_{22} - c_{28} = 0
\nonumber \\
c_{7} + c_{25} + c_{28} + c_{29} = 0
\label{c1}
\eea
and in the sector
$
{\cal D}{\cal D}d(x,y,z) u u F
$
\bea
c_{14} + c_{23} = b_{1}
\nonumber \\
c_{11} - c_{19} = b_{5} + f_{0}
\nonumber \\
c_{17} + c_{19} - c_{23} = b_{2}
\nonumber \\
- c_{10} + c_{20} = b_{6}
\nonumber \\
- c_{13} - c_{22} = b_{8}
\nonumber \\
- c_{16} - c_{20} + c_{22} = b_{5}
\nonumber \\
c_{10} - c_{14} + c_{26} = b_{3}
\nonumber \\
- c_{11} + c_{13} - c_{25} = b_{6}
\nonumber \\
c_{16} - c_{17} + c_{25} - c_{26} = b_{1}
\nonumber \\
- c_{15} + c_{24} = 0
\nonumber \\
- c_{12} - c_{21} = 0
\nonumber \\
- c_{18} + c_{21} + c_{24} = 0
\nonumber \\
c_{12} + c_{15} - c_{27} = 0
\nonumber \\
c_{18} + c_{27} = 0
\nonumber \\
- c_{1} + c_{2} = b_{7}
\nonumber \\
c_{1} - c_{5} + c_{8} = b_{4}
\nonumber \\
- c_{2} + c_{4} - c_{7} = b_{7}
\nonumber \\
- 2 c_{3} = f_{0}
\nonumber \\
c_{3} + c_{6} - c_{9} = 0
\label{c2}
\eea

From the equation (\ref{co2})
\be
D^{[\mu][\nu]\emptyset} 
+ {i \over 2} \sum_{j} a_{j,abc} b_{j}^{[\mu][\nu][\rho\sigma]}
W_{3,\rho\sigma} = 
(\bar{s}B_{2})^{[\mu][\nu]\emptyset}
+ D_{\delta}^{[\mu][\nu]\emptyset}
\nonumber
\ee 
it follows in the sector
$
{\cal D}{\cal D}d(x,y,z) u u \partial F
$
\bea
- c_{12} + d_{10} = 0
\nonumber \\
c_{12} + c_{21} - c_{30} = 0
\nonumber \\
- c_{15} + c_{28} = 0
\nonumber \\
c_{15} + c_{24} + c_{30} = 0
\nonumber \\
- c_{18} - c_{28} - d_{10} = 0
\nonumber \\
c_{18} + c_{27} = 0
\nonumber \\
- c_{29} + d_{5} = 0
\nonumber \\
c_{29} - d_{3} + d_{9} + d_{10} = 0
\nonumber \\
d_{3} - d_{8} - d_{10} = 0
\label{d1}
\eea
%\newpage
and in the sector
$
{\cal D}{\cal D}d(x,y,z) u u  F
$
%\newpage
\bea
c_{1} + c_{19} - d_{5} = 0
\nonumber \\
c_{10} - d_{1} - d_{6} = 0
\nonumber \\
c_{7} - d_{3} + d_{5} = 0
\nonumber \\
c_{16} - c_{27} - d_{4} + d_{6} = - f_{0}
\nonumber \\
c_{13} + c_{27} - d_{2} = - f_{0}
\nonumber \\
c_{4} - c_{19} + d_{3} = 0
\nonumber \\
- c_{18} + c_{25} + d_{8} = 0
\nonumber \\
- c_{9} + d_{1} + d_{7} = f_{0}
\nonumber \\
- c_{12} + d_{3} - d_{8} = 0
\nonumber \\
- c_{3} - c_{21} + d_{4} - d_{7} = 0
\nonumber \\
- c_{15} - c_{25} - d_{3} = 0
\nonumber \\
- c_{6} + c_{21} + d_{2} = f_{0}
\nonumber \\
- c_{1} + c_{18} + c_{22} - d_{9} = 0
\nonumber \\
c_{9} - c_{10} + d_{7} = 0
\nonumber \\
- c_{7} + c_{12} + d_{9} = 0
\nonumber \\
c_{3} - c_{16} - c_{24} - d_{7} = 0
\nonumber \\
- c_{4} + c_{15} - c_{22} = 0
\nonumber \\
c_{6} - c_{13} + c_{24} = 0
\nonumber \\
- c_{2} - c_{20} + d_{4} - d_{9} = a_{2}
\nonumber \\
- c_{11} + d_{2} + d_{8} = a_{1}
\nonumber \\
- c_{8} + d_{2} + d_{9} = - a_{3} - f_{0}
\nonumber \\
- c_{17} + c_{26} + d_{4} - d_{8} = a_{4}
\nonumber \\
- c_{5} + c_{20} + d_{1} = a_{10} - f_{0}
\nonumber \\
- c_{14} - c_{26} + d_{1} = a_{8}
\nonumber \\
c_{2} - c_{17} - c_{23} - d_{5} = a_{9} + f_{0}
\nonumber \\
- c_{8} + c_{11} - d_{5} = a_{7} - f_{0}
\nonumber \\
c_{5} - c_{14} + c_{23} = - a_{13} + f_{0}
\nonumber \\
- c_{16} - c_{18} + d_{1} - d_{2} = 0
\nonumber \\
c_{10} - c_{13} + c_{18} - d_{4} = 0
\nonumber \\
c_{12} - c_{15} + c_{16} + d_{4} = - f_{0}
\nonumber \\
2 c_{17} + 2 d_{3} = a_{6} - f_{0}
\nonumber \\
- c_{11} + c_{14} - c_{17} = a_{12}
\nonumber \\
- c_{25} - c_{27} - d_{5} - d_{9} = 0
\nonumber \\
- c_{19} - c_{22} + c_{27} + d_{8} = 0
\nonumber \\
c_{21} + c_{24} - c_{25} - d_{8} = 0
\nonumber \\
c_{20} + c_{23} - c_{26} - d_{6} = a_{11}
\nonumber \\
2 c_{26} + 2 d_{7} = a_{5}
\label{d2}
\eea

The systems (\ref{c1}) + (\ref{c2}) + (\ref{d1}) + (\ref{d2}) can be used to
obtain the parameters $c$ and $d$. No constraints are necessary on the physical
parameters of the system.

If we consider the expression
\be
D^{IJK} - (\bar{s}(B_{1} + B_{2})^{IJK}
\ee
we remain only with terms
$
\sim u F F
$
and
$
F F F
$.
\newpage
\subsection{The Causal Commutators for Ghost Number $1$}
We have the contribution
$
D_{1} \sim u v v, D_{2} \sim u v F
$
and
$
D_{3} \sim u F F
$:
\bea
D_{3}^{\emptyset\emptyset[\mu]}(x,y,z) = i f^{(0)}_{abc} 
[ {\cal D}_{1}^{\nu}d(x,y,z) F_{a\nu\rho}(x) F_{b}^{\mu\rho}(y) u_{c}(z)
\nonumber \\
- {\cal D}_{1}^{\mu}d(x,y,z) F_{a\rho\sigma}(x) F_{b}^{\rho\sigma}(y) u_{c}(z)
+ {\cal D}_{1}^{\nu}d(x,y,z) F_{a\nu\rho}(x) u_{b}(y) F_{c}^{\mu\rho}(z)]
\eea
%\newpage
\subsection{The Generic Form of the Coboundaries}
\bea
B_{3}^{\emptyset\emptyset[\mu\nu]}(x,y,z) = \sum \hat{c}_{j,abc} 
\hat{B}_{j}^{\emptyset\emptyset[\mu\nu]}(x,y,z)
\nonumber \\
B_{3}^{[\mu][\nu]\emptyset}(x,y,z) = \sum \hat{d}_{j,abc} 
\hat{B}_{j}^{[\mu][\nu]\emptyset}(x,y,z)
\eea
where
\bea
\hat{B}_{1}^{\emptyset\emptyset[\mu\nu]}(x,y,z) 
= d(x,y,z) \eta_{\rho\sigma} F_{a}^{\mu\rho}(x) F_{b}^{\nu\sigma}(y) u_{c}(z)
- (\mu \leftrightarrow \nu)
\nonumber \\
\hat{B}_{2}^{\emptyset\emptyset[\mu\nu]}(x,y,z) = d(x,y,z) \eta_{\rho\sigma} 
[ u_{a}(x) F_{b}^{\mu\rho}(y) F_{c}^{\nu\sigma}(z) 
- F_{a}^{\mu\rho}(x) u_{b}(y) F_{c}^{\nu\sigma}(z) ]
- (\mu \leftrightarrow \nu)
\eea
and
\bea
\hat{B}_{1}^{[\mu][\nu]\emptyset}(x,y,z) 
= d(x,y,z) \eta_{\rho\sigma} F_{a}^{\mu\rho}(x) F_{b}^{\nu\sigma}(y) u_{c}(z)
\nonumber \\
\hat{B}_{2}^{[\mu][\nu]\emptyset}(x,y,z) 
= d(x,y,z) \eta_{\rho\sigma} F_{a}^{\nu\rho}(x) F_{b}^{\mu\sigma}(y) u_{c}(z)
\nonumber \\
\hat{B}_{3}^{[\mu][\nu]\emptyset}(x,y,z) = d(x,y,z) \eta_{\rho\sigma} 
[ u_{a}(x) F_{b}^{\mu\rho}(y) F_{c}^{\nu\sigma}(z) 
+ F_{a}^{\nu\rho}(x) u_{b}(y) F_{c}^{\mu\sigma}(z) ]
\nonumber \\
\hat{B}_{4}^{[\mu][\nu]\emptyset}(x,y,z) = d(x,y,z) \eta_{\rho\sigma} 
[ u_{a}(x) F_{b}^{\nu\rho}(y) F_{c}^{\mu\sigma}(z) 
+ F_{a}^{\mu\rho}(x) u_{b}(y) F_{c}^{\nu\sigma}(z) ]
\nonumber \\
\hat{B}_{5}^{[\mu][\nu]\emptyset}(x,y,z) = d(x,y,z) \eta^{\mu\nu} 
F_{a}^{\rho\sigma}(y) F_{b\rho\sigma}(z) u_{c}(z)
\nonumber \\
\hat{B}_{6}^{[\mu][\nu]\emptyset}(x,y,z) = d(x,y,z) \eta^{\mu\nu}
[ u_{a}(x) F_{a}^{\rho\sigma}(y) F_{b\rho\sigma}(z) 
+ F_{a}^{\rho\sigma}(x) u_{b}(y) F_{c\rho\sigma}(z) ]
\eea

We consider the equation
\be
D_{3}^{\emptyset\emptyset[\mu]}
+ \tilde{D}_{3}^{\emptyset\emptyset[\mu]} =
(\bar{s}B_{3})^{\emptyset\emptyset[\mu]} 
\ee
where
$
\tilde{D}_{3}
$
contains the contributions coming from
$
- (\bar{s}(B_{1} + B_{2}))^{IJK}
$;
we obtain in the sector
$
{\cal D}d(X) F F u
$
the following equations
\bea
\hat{c}_{1} + \hat{d}_{3} = - c_{17} - c_{23} + f_{0}
\nonumber \\
- \hat{c}_{1} + \hat{d}_{4} = - c_{14} - c_{26}
\nonumber \\
\hat{c}_{2} + \hat{d}_{3} = - c_{25}
\nonumber \\
- \hat{c}_{2} + \hat{d}_{2} = - c_{22}
\nonumber \\
- \hat{c}_{2} + \hat{d}_{4} = - c_{16} + f_{0}
\nonumber \\
\hat{c}_{2} + \hat{d}_{1} = - c_{13}
\nonumber \\
- \hat{d}_{1} - \hat{d}_{3} = - c_{10}
\nonumber \\
- \hat{d}_{2} - \hat{d}_{3} = - c_{19}
\nonumber \\
- \hat{d}_{3} - \hat{d}_{4} = - c_{11} - c_{20}
\nonumber \\
\hat{d}_{5} = - {1\over 2} c_{4}
\nonumber \\
\hat{d}_{6} = - {1\over 2} c_{7}
\nonumber \\
- \hat{d}_{5} -\hat{d}_{6}  = - {1\over 2} c_{1}
\nonumber \\
- 2 \hat{d}_{6} = - c_{2}
\nonumber \\
\hat{d}_{6} = - {1\over 2} (c_{5} + c_{8}) - f_{0}
\nonumber \\
- {1\over 2} \hat{c}_{1} = - {1\over 2} c_{28}
\nonumber \\
- {1\over 2} \hat{d}_{2} - \hat{d}_{5} = 0
\nonumber \\
- {1\over 2} \hat{d}_{3} - \hat{d}_{6} = - {1\over 2} c_{29}
\eea

We can solve this system {\it iff} we impose
\be
f^{(3)}_{abc} = 6 f^{(0)}_{abc}
\label{fff2}
\ee
and if we combine with (\ref{fff1}) we get also
\be
f^{(4)}_{abc} = 8 f^{(0)}_{abc}
\label{fff3}
\ee

One can show also by direct computation that the contribution
$
D_{4}
$
bilinear in the scalar fields 
$
\Phi_{a}
$
does not produces new restriction.
\newpage
\subsection{The Causal Commutators in the Dirac Sector}
We define
\bea
t^{(1)}_{a\epsilon} \equiv \sum_{b} g_{ab} t^{\epsilon}_{b} \qquad
t^{(2)}_{a\epsilon} \equiv 
\sum_{b} t^{\epsilon}_{b} t^{\epsilon}_{a} t^{\epsilon}_{b}
\nonumber \\
t^{(3)}_{a\epsilon} \equiv 
\sum_{b} s^{- \epsilon}_{b} t^{\epsilon}_{a} s^{\epsilon}_{b} \quad
t^{(4)}_{a\epsilon} \equiv - 
i \sum_{b,c}  f^{\prime}_{bca} s^{- \epsilon}_{b} s^{\epsilon}_{c}
\eea
and we have the following Dirac contribution in ghost number $1$:
\bea
D_{5}(T(x), T(y); T^{\mu}(z)) = 
\nonumber \\
{i \over 2} {\cal D}_{1}^{\nu}{\cal D}_{2}^{\rho}d(x,y,z) u_{a}(x) 
[ \bar{\Psi}(y) t^{(1)}_{a\epsilon} \otimes
\gamma_{\rho} \gamma_{\nu} \gamma^{\mu} \gamma_{\epsilon} \Psi(z)
+ \bar{\Psi}(z) t^{(1)}_{a\epsilon} \otimes 
\gamma^{\mu} \gamma_{\nu} \gamma_{\rho} \gamma_{\epsilon} \Psi(y) ]
\nonumber \\
+ 2 i {\cal D}_{1}^{\nu}{\cal D}_{2}^{\rho}d(x,y,z) u_{a}(z) 
\bar{\Psi}(x) t^{(2)}_{a\epsilon} \otimes
\gamma_{\nu} \gamma_{\rho} \gamma^{\mu} \gamma_{\epsilon} \Psi(y)
\nonumber \\
- 4 i {\cal D}_{1}^{\nu}{\cal D}_{2}^{\mu}d(x,y,z) u_{a}(z) 
\bar{\Psi}(x) t^{(2)}_{a\epsilon} \otimes \gamma_{\nu} \gamma_{\epsilon} \Psi(y)
\nonumber \\
- i {\cal D}_{1}^{\nu}{\cal D}_{2}^{\rho}d(x,y,z) u_{a}(z) 
\bar{\Psi}(x) t^{(3)}_{a\epsilon} \otimes 
\gamma_{\rho} \gamma^{\mu} \gamma_{\nu} \gamma_{\epsilon} \Psi(y)
\nonumber \\
+ {i \over 2} {\cal D}_{1}^{\nu}{\cal D}_{3}^{\rho}d(x,y,z) u_{a}(z) 
\bar{\Psi}(x) t^{(1)}_{a\epsilon} \otimes
\gamma_{\nu} \gamma_{\rho} \gamma^{\mu} \gamma_{\epsilon} \Psi(y)
\nonumber \\
+ i {\cal D}_{1}^{\mu}{\cal D}_{3}^{\nu}d(x,y,z) u_{a}(z) 
\bar{\Psi}(x) t^{(1)}_{a\epsilon} \otimes \gamma_{\nu} \gamma_{\epsilon} \Psi(y)
\nonumber \\
+ {i \over 2} {\cal D}_{2}^{\nu}{\cal D}_{3}^{\rho}d(x,y,z) u_{a}(z) 
\bar{\Psi}(x) t^{(1)}_{a\epsilon} \otimes
\gamma^{\mu} \gamma_{\rho} \gamma_{\nu} \gamma_{\epsilon} \Psi(y)
\nonumber \\
+ i {\cal D}_{2}^{\mu}{\cal D}_{3}^{\nu}d(x,y,z) u_{a}(z) 
\bar{\Psi}(x) t^{(1)}_{a\epsilon} \otimes \gamma_{\nu} \gamma_{\epsilon} \Psi(y)
\nonumber \\
+ i {\cal D}_{1}^{\mu}{\cal D}_{3}^{\nu}d(x,y,z) u_{a}(z) 
\bar{\Psi}(x) t^{(4)}_{a\epsilon} \otimes \gamma_{\nu} \gamma_{\epsilon} \Psi(y)
\nonumber \\
+ i {\cal D}_{2}^{\mu}{\cal D}_{3}^{\nu}d(x,y,z) u_{a}(z) 
\bar{\Psi}(x) t^{(4)}_{a\epsilon} \otimes \gamma_{\nu} \gamma_{\epsilon} \Psi(y)
\eea
\newpage
\subsection{The Generic Form of the Coboundaries}
We impose the coboundary condition
\be
D_{5}^{\emptyset\emptyset[\mu]}(x,y,z)_{0} = 
(\bar{s}B_{5})^{\emptyset\emptyset[\mu]}(x,y,z)
+ D_{5\delta}^{\emptyset\emptyset[\mu]}(x,y,z)
\label{co3}
\ee
and from the various restrictions we have the generic forms:
\bea
B_{5}^{[\mu][\nu]}(x,y,z) = 
\nonumber \\
{\cal D}_{1}^{\mu}d(X) u_{a}(x) 
\bar{\Psi}(y) F^{(1)}_{a\epsilon} \otimes \gamma^{\nu} \gamma_{\epsilon} \Psi(z)
%\nonumber \\
+ {\cal D}_{2}^{\nu}d(X) u_{a}(y) 
\bar{\Psi}(x) F^{(1)}_{a\epsilon} \otimes \gamma^{\mu} \gamma_{\epsilon} \Psi(z)
\nonumber \\
+ {\cal D}_{1}^{\nu}d(X) u_{a}(x) 
\bar{\Psi}(y) F^{(2)}_{a\epsilon} \otimes \gamma^{\mu} \gamma_{\epsilon} \Psi(z)
%\nonumber \\
+ {\cal D}_{2}^{\mu}d(X) u_{a}(y) 
\bar{\Psi}(x) F^{(2)}_{a\epsilon} \otimes \gamma^{\nu} \gamma_{\epsilon} \Psi(z)
\nonumber \\
+ {\cal D}_{2}^{\mu}d(X) u_{a}(x) 
\bar{\Psi}(y) F^{(3)}_{a\epsilon} \otimes \gamma^{\nu} \gamma_{\epsilon} \Psi(z)
%\nonumber \\
+ {\cal D}_{1}^{\nu}d(X) u_{a}(y) 
\bar{\Psi}(x) F^{(3)}_{a\epsilon} \otimes \gamma^{\mu} \gamma_{\epsilon} \Psi(z)
\nonumber \\
+ {\cal D}_{2}^{\nu}d(X) u_{a}(x) 
\bar{\Psi}(y) F^{(4)}_{a\epsilon} \otimes \gamma^{\mu} \gamma_{\epsilon} \Psi(z)
%\nonumber \\
+ {\cal D}_{1}^{\mu}d(X) u_{a}(y) 
\bar{\Psi}(x) F^{(4)}_{a\epsilon} \otimes \gamma^{\nu} \gamma_{\epsilon} \Psi(z)
\nonumber \\
+ {\cal D}_{3}^{\mu}d(X) u_{a}(x) 
\bar{\Psi}(y) F^{(5)}_{a\epsilon} \otimes \gamma^{\nu} \gamma_{\epsilon} \Psi(z)
%\nonumber \\
+ {\cal D}_{3}^{\nu}d(X) u_{a}(y) 
\bar{\Psi}(x) F^{(5)}_{a\epsilon} \otimes \gamma^{\mu} \gamma_{\epsilon} \Psi(z)
\nonumber \\
+ {\cal D}_{3}^{\nu}d(X) u_{a}(x) 
\bar{\Psi}(y) F^{(6)}_{a\epsilon} \otimes \gamma^{\mu} \gamma_{\epsilon} \Psi(z)
%\nonumber \\
+ {\cal D}_{3}^{\mu}d(X) u_{a}(y) 
\bar{\Psi}(x) F^{(6)}_{a\epsilon} \otimes \gamma^{\nu} \gamma_{\epsilon} \Psi(z)
\nonumber \\
+ \eta^{\mu\nu} [ {\cal D}_{1}^{\rho}d(X) u_{a}(x) 
\bar{\Psi}(y) F^{(7)}_{a\epsilon} \otimes \gamma_{\rho} \gamma_{\epsilon} \Psi(z)
%\nonumber \\
+ {\cal D}_{2}^{\rho}d(X) u_{a}(y) 
\bar{\Psi}(x) F^{(7)}_{a\epsilon} \otimes \gamma_{\rho} \gamma_{\epsilon} \Psi(z) ]
\nonumber \\
+ \eta^{\mu\nu} [ {\cal D}_{2}^{\rho}d(X) u_{a}(x) 
\bar{\Psi}(y) F^{(8)}_{a\epsilon} \otimes \gamma_{\rho} \gamma_{\epsilon} \Psi(z)
%\nonumber \\
+ {\cal D}_{1}^{\rho}d(X) u_{a}(y) 
\bar{\Psi}(x) F^{(8)}_{a\epsilon} \otimes \gamma_{\rho} \gamma_{\epsilon} \Psi(z) ]
\nonumber \\
+ \eta^{\mu\nu} [ {\cal D}_{3}^{\rho}d(X) u_{a}(x) 
\bar{\Psi}(y) F^{(9)}_{a\epsilon} \otimes \gamma_{\rho} \gamma_{\epsilon} \Psi(z)
%\nonumber \\
+ {\cal D}_{3}^{\rho}d(X) u_{a}(y) 
\bar{\Psi}(x) F^{(9)}_{a\epsilon} \otimes \gamma_{\rho} \gamma_{\epsilon} \Psi(z) ]
\nonumber \\
+ {\cal D}_{1}^{\mu}d u_{a}(x) 
\bar{\Psi}(z) F^{(10)}_{a\epsilon} \otimes \gamma^{\nu} \gamma_{\epsilon} \Psi(y)
%\nonumber \\
+ {\cal D}_{2}^{\nu}d(X) u_{a}(y) 
\bar{\Psi}(z) F^{(10)}_{a\epsilon} \otimes \gamma^{\mu} \gamma_{\epsilon} \Psi(x)
\nonumber \\
+ {\cal D}_{1}^{\nu}d(X) u_{a}(x) 
\bar{\Psi}(z) F^{(11)}_{a\epsilon} \otimes \gamma^{\mu} \gamma_{\epsilon} \Psi(y)
%\nonumber \\
+ {\cal D}_{2}^{\mu}d(X) u_{a}(y) 
\bar{\Psi}(z) F^{(11)}_{a\epsilon} \otimes \gamma^{\nu} \gamma_{\epsilon} \Psi(x)
\nonumber \\
+ {\cal D}_{2}^{\mu}d(X) u_{a}(x) 
\bar{\Psi}(z) F^{(12)}_{a\epsilon} \otimes \gamma^{\nu} \gamma_{\epsilon} \Psi(y)
%\nonumber \\
+ {\cal D}_{1}^{\nu}d(X) u_{a}(y) 
\bar{\Psi}(z) F^{(12)}_{a\epsilon} \otimes \gamma^{\mu} \gamma_{\epsilon} \Psi(x)
\nonumber \\
+ {\cal D}_{2}^{\nu}d(X) u_{a}(x) 
\bar{\Psi}(z) F^{(13)}_{a\epsilon} \otimes \gamma^{\mu} \gamma_{\epsilon} \Psi(y)
%\nonumber \\
+ {\cal D}_{1}^{\mu}d(X) u_{a}(y) 
\bar{\Psi}(z) F^{(13)}_{a\epsilon} \otimes \gamma^{\nu} \gamma_{\epsilon} \Psi(x)
\nonumber \\
+ {\cal D}_{3}^{\mu}d(X) u_{a}(x) 
\bar{\Psi}(z) F^{(14)}_{a\epsilon} \otimes \gamma^{\nu} \gamma_{\epsilon} \Psi(y)
%\nonumber \\
+ {\cal D}_{3}^{\nu}d(X) u_{a}(y) 
\bar{\Psi}(z) F^{(14)}_{a\epsilon} \otimes \gamma^{\mu} \gamma_{\epsilon} \Psi(x)
\nonumber \\
+ {\cal D}_{3}^{\nu}d(X) u_{a}(x) 
\bar{\Psi}(z) F^{(15)}_{a\epsilon} \otimes \gamma^{\mu} \gamma_{\epsilon} \Psi(y)
%\nonumber \\
+ {\cal D}_{3}^{\mu}d(X) u_{a}(y) 
\bar{\Psi}(z) F^{(15)}_{a\epsilon} \otimes \gamma^{\nu} \gamma_{\epsilon} \Psi(x)
\nonumber \\
+ \eta^{\mu\nu} [ {\cal D}_{1}^{\rho}d(X) u_{a}(x) 
\bar{\Psi}(z) F^{(16)}_{a\epsilon} \otimes \gamma_{\rho} \gamma_{\epsilon} \Psi(y)
%\nonumber \\
+ {\cal D}_{2}^{\rho}d(X) u_{a}(y) 
\bar{\Psi}(z) F^{(16)}_{a\epsilon} \otimes \gamma_{\rho} \gamma_{\epsilon} \Psi(x) ]
\nonumber \\
+ \eta^{\mu\nu} [ {\cal D}_{2}^{\rho}d(X) u_{a}(x) 
\bar{\Psi}(z) F^{(17)}_{a\epsilon} \otimes \gamma_{\rho} \gamma_{\epsilon} \Psi(y)
%\nonumber \\
+ {\cal D}_{1}^{\rho}d(X) u_{a}(y)
\bar{\Psi}(z) F^{(17)}_{a\epsilon} \otimes \gamma_{\rho} \gamma_{\epsilon} \Psi(x) ]
\nonumber \\
+ \eta^{\mu\nu} [ {\cal D}_{3}^{\rho}d(X) u_{a}(x) 
\bar{\Psi}(z) F^{(18)}_{a\epsilon} \otimes \gamma_{\rho} \gamma_{\epsilon} \Psi(y)
%\nonumber \\
+ {\cal D}_{3}^{\rho}d(X) u_{a}(y) 
\bar{\Psi}(z) F^{(18)}_{a\epsilon} \otimes \gamma_{\rho} \gamma_{\epsilon} \Psi(x) ]
\nonumber \\
+  {\cal D}_{1}^{\mu}d(X) u_{a}(z) 
\bar{\Psi}(x) F^{(19)}_{a\epsilon} \otimes \gamma^{\nu} \gamma_{\epsilon} \Psi(y)
%\nonumber \\
+ {\cal D}_{2}^{\nu}d u_{a}(z) 
\bar{\Psi}(y) F^{(19)}_{a\epsilon} \otimes \gamma^{\mu} \gamma_{\epsilon} \Psi(x)
\nonumber \\
+ {\cal D}_{1}^{\nu}d(X) u_{a}(z) 
\bar{\Psi}(x) F^{(20)}_{a\epsilon} \otimes \gamma^{\mu} \gamma_{\epsilon} \Psi(y)
%\nonumber \\
+ {\cal D}_{2}^{\mu}d(X) u_{a}(z) 
\bar{\Psi}(y) F^{(20)}_{a\epsilon} \otimes \gamma^{\nu} \gamma_{\epsilon} \Psi(x)
\nonumber \\
+ {\cal D}_{2}^{\mu}d(X) u_{a}(z) 
\bar{\Psi}(x) F^{(21)}_{a\epsilon} \otimes \gamma^{\nu} \gamma_{\epsilon} \Psi(y)
%\nonumber \\
+ {\cal D}_{1}^{\nu}d u_{a}(z) 
\bar{\Psi}(y) F^{(21)}_{a\epsilon} \otimes \gamma^{\mu} \gamma_{\epsilon} \Psi(x)
\nonumber \\
+ {\cal D}_{2}^{\nu}d(X) u_{a}(z) 
\bar{\Psi}(x) F^{(22)}_{a\epsilon} \otimes \gamma^{\mu} \gamma_{\epsilon} \Psi(y)
%\nonumber \\
+ {\cal D}_{1}^{\mu}d(X) u_{a}(z) 
\bar{\Psi}(y) F^{(22)}_{a\epsilon} \otimes \gamma^{\nu} \gamma_{\epsilon} \Psi(x)
\nonumber \\
+ {\cal D}_{3}^{\mu}d(X) u_{a}(z) 
\bar{\Psi}(x) F^{(23)}_{a\epsilon} \otimes \gamma^{\nu} \gamma_{\epsilon} \Psi(y)
%\nonumber \\
+ {\cal D}_{3}^{\nu}d(X) u_{a}(z) 
\bar{\Psi}(y) F^{(23)}_{a\epsilon} \otimes \gamma^{\mu} \gamma_{\epsilon} \Psi(x)
\nonumber \\
+ {\cal D}_{3}^{\nu}d(X) u_{a}(z) 
\bar{\Psi}(x) F^{(24)}_{a\epsilon} \otimes \gamma^{\mu} \gamma_{\epsilon} \Psi(y)
%\nonumber \\
+ {\cal D}_{3}^{\mu}d(X) u_{a}(z) 
\bar{\Psi}(y) F^{(24)}_{a\epsilon} \otimes \gamma^{\nu} \gamma_{\epsilon} \Psi(x)
\nonumber \\
+ \eta^{\mu\nu} [ {\cal D}_{1}^{\rho}d(X) u_{a}(z) 
\bar{\Psi}(x) F^{(25)}_{a\epsilon} \otimes \gamma_{\rho} \gamma_{\epsilon} \Psi(y)
%\nonumber \\
+ {\cal D}_{2}^{\rho}d(X) u_{a}(z) 
\bar{\Psi}(y) F^{(25)}_{a\epsilon} \otimes \gamma_{\rho} \gamma_{\epsilon} \Psi(x) ]
\nonumber \\
+ \eta^{\mu\nu} [ {\cal D}_{2}^{\rho}d(X) u_{a}(z) 
\bar{\Psi}(x) F^{(26)}_{a\epsilon} \otimes \gamma_{\rho} \gamma_{\epsilon} \Psi(y)
%\nonumber \\
+ {\cal D}_{1}^{\rho}d(X) u_{a}(z) 
\bar{\Psi}(y) F^{(26)}_{a\epsilon} \otimes \gamma_{\rho} \gamma_{\epsilon} \Psi(x) ]
\nonumber \\
+ \eta^{\mu\nu} [ {\cal D}_{3}^{\rho}d(X) u_{a}(z) 
\bar{\Psi}(x) F^{(27)}_{a\epsilon} \otimes \gamma_{\rho} \gamma_{\epsilon} \Psi(y)
%\nonumber \\
+ {\cal D}_{3}^{\rho}d(X) u_{a}(z) 
\bar{\Psi}(y) F^{(27)}_{a\epsilon} \otimes \gamma_{\rho} \gamma_{\epsilon} \Psi(x) ]
\nonumber\\
+ d(X) [ u_{a}(x) \partial^{\mu}\bar{\Psi}(y) F^{(28)}_{a\epsilon} \otimes \gamma^{\nu}
\gamma_{\epsilon} \Psi(z)
%\nonumber \\
- u_{a}(y) \partial^{\nu}\bar{\Psi}(x) F^{(28)}_{a\epsilon} \otimes
\gamma^{\mu}] \gamma_{\epsilon} \Psi(z)
\nonumber \\
+ d(X) [ u_{a}(x) \partial^{\nu}\bar{\Psi}(y) F^{(29)}_{a\epsilon} \otimes \gamma^{\mu}
\gamma_{\epsilon} \Psi(z)
%\nonumber \\
- u_{a}(y) \partial^{\mu}\bar{\Psi}(x) F^{(29)}_{a\epsilon} \otimes
\gamma^{\nu} \gamma_{\epsilon} \Psi(z) ]
\nonumber \\
+ d(X) [ u_{a}(x) \bar{\Psi}(y) F^{(30)}_{a\epsilon} \otimes \gamma^{\nu}
\gamma_{\epsilon} \partial^{\mu}\Psi(z)
%\nonumber \\
- u_{a}(y) \bar{\Psi}(x) F^{(30)}_{a\epsilon} \otimes \gamma^{\mu}
\gamma_{\epsilon} \partial^{\nu}\Psi(z) ]
\nonumber \\
+ d(X) [ u_{a}(x) \bar{\Psi}(y) F^{(31)}_{a\epsilon} \otimes \gamma^{\mu}
\gamma_{\epsilon} \partial^{\nu}\Psi(z)
%\nonumber \\
- u_{a}(y) \bar{\Psi}(x) F^{(31)}_{a\epsilon} \otimes \gamma^{\nu}
\gamma_{\epsilon} \partial^{\mu}\Psi(z) ]
\nonumber \\
+ d(X) [ u_{a}(x) \partial^{\mu}\bar{\Psi}(z) F^{(32)}_{a\epsilon} \otimes \gamma^{\nu}
\gamma_{\epsilon} \Psi(y)
%\nonumber \\
- u_{a}(y) \partial^{\nu}\bar{\Psi}(z) F^{(32)}_{a\epsilon} \otimes \gamma^{\mu}
\gamma_{\epsilon} \Psi(x) ]
\nonumber \\
+ d(X) [ u_{a}(x) \partial^{\nu}\bar{\Psi}(z) F^{(33)}_{a\epsilon} \otimes \gamma^{\mu}
\gamma_{\epsilon} \Psi(y)
%\nonumber \\
- u_{a}(y) \partial^{\mu}\bar{\Psi}(z) F^{(33)}_{a\epsilon} \otimes \gamma^{\nu}
\gamma_{\epsilon} \Psi(x) ]
\nonumber \\
+ d(X) [ u_{a}(x) \bar{\Psi}(z) F^{(34)}_{a\epsilon} \otimes \gamma^{\nu}
\gamma_{\epsilon} \partial^{\mu}\Psi(z)
%\nonumber \\
- u_{a}(y) \bar{\Psi}(z) F^{(34)}_{a\epsilon} \otimes \gamma^{\mu}
\gamma_{\epsilon} \partial^{\nu}\Psi(z) ]
\nonumber \\
+ d(X) [ u_{a}(x) \bar{\Psi}(z) F^{(35)}_{a\epsilon} \otimes \gamma^{\mu}
\gamma_{\epsilon} \partial^{\nu}\Psi(z) ]
%\nonumber \\
- u_{a}(y) \bar{\Psi}(z) F^{(35)}_{a\epsilon} \otimes \gamma^{\nu}
\gamma_{\epsilon} \partial^{\mu}\Psi(z) ]
\nonumber \\
+ d(X) u_{a}(z) 
[ \partial^{\mu}\bar{\Psi}(x) F^{(36)}_{a\epsilon} \otimes \gamma^{\nu}
\gamma_{\epsilon} \Psi(y)
%\nonumber \\
- \partial^{\nu}\bar{\Psi}(y) F^{(36)}_{a\epsilon} \otimes \gamma^{\mu}
\gamma_{\epsilon} \Psi(x) ]
\nonumber \\
+ d(X) u_{a}(z) 
[ \partial^{\nu}\bar{\Psi}(x) F^{(37)}_{a\epsilon} \otimes \gamma^{\mu}
\gamma_{\epsilon} \Psi(y)
%\nonumber \\
- \partial^{\mu}\bar{\Psi}(y) F^{(37)}_{a\epsilon} \otimes \gamma^{\nu}
\gamma_{\epsilon} \Psi(x) ]
\nonumber \\
+ d(X) u_{a}(z) 
[ \bar{\Psi}(x) F^{(38)}_{a\epsilon} \otimes \gamma^{\nu}
\gamma_{\epsilon} \partial^{\mu}\Psi(y)
%\nonumber \\
- \bar{\Psi}(y) F^{(38)}_{a\epsilon} \otimes \gamma^{\mu}
\gamma_{\epsilon} \partial^{\nu}\Psi(x) ]
\nonumber \\
+ d(X) u_{a}(z) 
[ \bar{\Psi}(x) F^{(39)}_{a\epsilon} \otimes \gamma^{\mu}
\gamma_{\epsilon} \partial^{\nu}\Psi(y)
%\nonumber \\
- \bar{\Psi}(y) F^{(39)}_{a\epsilon} \otimes \gamma^{\nu}
\gamma_{\epsilon} \partial^{\mu}\Psi(x) ]
\nonumber\\
+ {\cal D}_{1}^{\rho}d(X) u_{a}(x) 
\bar{\Psi}(y) F^{(40)}_{a\epsilon} \otimes \gamma^{\mu} \gamma^{\nu} 
\gamma_{\rho} \gamma_{\epsilon} \Psi(z)
%\nonumber \\
+ {\cal D}_{2}^{\rho}d(X)
u_{a}(y) \bar{\Psi}(x) F^{(40)}_{a\epsilon} \otimes 
\gamma^{\nu} \gamma^{\mu} \gamma_{\rho} \gamma_{\epsilon} \Psi(z)
\nonumber \\
+ {\cal D}_{2}^{\rho}d(X)
u_{a}(x) \bar{\Psi}(y) F^{(41)}_{a\epsilon} \otimes 
\gamma^{\mu} \gamma^{\nu} \gamma_{\rho} \gamma_{\epsilon} \Psi(z)
%\nonumber \\
+ {\cal D}_{1}^{\rho}d(X)
u_{a}(y) \bar{\Psi}(x) F^{(41)}_{a\epsilon} \otimes 
\gamma^{\nu} \gamma^{\mu} \gamma_{\rho} \gamma_{\epsilon} \Psi(z)
\nonumber \\
+ {\cal D}_{3}^{\rho}d(X)
[ u_{a}(x) \bar{\Psi}(y) F^{(42)}_{a\epsilon} \otimes 
\gamma^{\mu} \gamma^{\nu} \gamma_{\rho} \gamma_{\epsilon} \Psi(z)
%\nonumber \\
+ u_{a}(y) \bar{\Psi}(x) F^{(42)}_{a\epsilon} \otimes 
\gamma^{\nu} \gamma^{\mu} \gamma_{\rho} \gamma_{\epsilon} \Psi(z) ]
\nonumber \\
+ {\cal D}_{1}^{\rho}d(X)
u_{a}(x) \bar{\Psi}(z) F^{(43)}_{a\epsilon} \otimes 
\gamma^{\mu} \gamma^{\nu} \gamma_{\rho} \gamma_{\epsilon} \Psi(y)
%\nonumber \\
+ {\cal D}_{2}^{\rho}d(X)
u_{a}(y) \bar{\Psi}(z) F^{(43)}_{a\epsilon} \otimes 
\gamma^{\nu} \gamma^{\mu} \gamma_{\rho} \gamma_{\epsilon} \Psi(x)
\nonumber \\
+ {\cal D}_{2}^{\rho}d(X)
u_{a}(x) \bar{\Psi}(z) F^{(44)}_{a\epsilon} \otimes 
\gamma^{\mu} \gamma^{\nu} \gamma_{\rho} \gamma_{\epsilon} \Psi(y)
%\nonumber \\
+ {\cal D}_{1}^{\rho}d(X)
u_{a}(y) \bar{\Psi}(z) F^{(44)}_{a\epsilon} \otimes 
\gamma^{\nu}\gamma^{\mu} \gamma_{\rho} \gamma_{\epsilon} \Psi(x)
\nonumber \\
+ {\cal D}_{3}^{\rho}d(X)
[ u_{a}(x) \bar{\Psi}(z) F^{(45)}_{a\epsilon} \otimes 
\gamma^{\mu} \gamma^{\nu} \gamma_{\rho} \gamma_{\epsilon} \Psi(y)
%\nonumber \\
+ u_{a}(y) \bar{\Psi}(z) F^{(45)}_{a\epsilon} \otimes 
\gamma^{\nu} \gamma^{\mu} \gamma_{\rho} \gamma_{\epsilon} \Psi(x) ]
\nonumber \\
+ {\cal D}_{1}^{\rho}d(X)
u_{a}(z) \bar{\Psi}(x) F^{(46)}_{a\epsilon} \otimes 
\gamma^{\mu} \gamma^{\nu} \gamma_{\rho} \gamma_{\epsilon} \Psi(y)
%\nonumber \\
+ {\cal D}_{2}^{\rho}d(X)
u_{a}(z) \bar{\Psi}(y) F^{(46)}_{a\epsilon} \otimes 
\gamma^{\nu} \gamma^{\mu} \gamma_{\rho} \gamma_{\epsilon} \Psi(x)
\nonumber \\
+ {\cal D}_{2}^{\rho}d(X)
u_{a}(z) \bar{\Psi}(x) F^{(47)}_{a\epsilon} \otimes 
\gamma^{\mu} \gamma^{\nu} \gamma_{\rho} \gamma_{\epsilon} \Psi(y)
%\nonumber \\
+ {\cal D}_{1}^{\rho}d(X)
u_{a}(z) \bar{\Psi}(y) F^{(47)}_{a\epsilon} \otimes 
\gamma^{\nu}\gamma^{\mu} \gamma_{\rho} \gamma_{\epsilon} \Psi(x)
\nonumber \\
+ {\cal D}_{3}^{\rho}d(X) u_{a}(z)
[ \bar{\Psi}(x) F^{(48)}_{a\epsilon} \otimes 
\gamma^{\mu} \gamma^{\nu} \gamma_{\rho} \gamma_{\epsilon} \Psi(y)
%\nonumber \\
+ \bar{\Psi}(y) F^{(48)}_{a\epsilon} \otimes 
\gamma^{\nu} \gamma^{\mu} \gamma_{\rho} \gamma_{\epsilon} \Psi(x) ]
\eea
%\newpage
and
\bea
B_{5}^{\emptyset\emptyset[\mu\nu]}(x,y,z) = 
\nonumber \\
\{ {\cal D}_{1}^{\mu}d(x,y,z) u_{a}(x) 
\bar{\Psi}(y) G^{(1)}_{a\epsilon} \otimes \gamma^{\nu} \gamma_{\epsilon} \Psi(z)
%\nonumber \\
- {\cal D}_{2}^{\mu}d(x,y,z) u_{a}(y) 
\bar{\Psi}(x) G^{(1)}_{a\epsilon} \otimes \gamma^{\mu} \gamma_{\epsilon} \Psi(z)
\nonumber \\
+ {\cal D}_{2}^{\mu}d(x,y,z) u_{a}(x) 
\bar{\Psi}(y) G^{(2)}_{a\epsilon} \otimes \gamma^{\nu} \gamma_{\epsilon} \Psi(z)
%\nonumber \\
- {\cal D}_{1}^{\mu}d(x,y,z) u_{a}(y) 
\bar{\Psi}(x) G^{(2)}_{a\epsilon} \otimes \gamma^{\nu} \gamma_{\epsilon} \Psi(z) 
\nonumber \\
+ {\cal D}_{3}^{\mu}d(x,y,z) u_{a}(x) 
\bar{\Psi}(y) G^{(3)}_{a\epsilon} \otimes \gamma^{\nu} \gamma_{\epsilon} \Psi(z)
%\nonumber \\
- {\cal D}_{3}^{\mu}d(x,y,z) u_{a}(y) 
\bar{\Psi}(x) G^{(3)}_{a\epsilon} \otimes \gamma^{\mu} \gamma_{\epsilon} \Psi(z) 
\nonumber \\
+ {\cal D}_{1}^{\mu}d(x,y,z) u_{a}(x) 
\bar{\Psi}(z) G^{(4)}_{a\epsilon} \otimes \gamma^{\mu} \gamma_{\epsilon} \Psi(y)
%\nonumber \\
- {\cal D}_{2}^{\mu}d(x,y,z) u_{a}(y) 
\bar{\Psi}(z) G^{(4)}_{a\epsilon} \otimes \gamma^{\nu} \gamma_{\epsilon} \Psi(x) 
\nonumber \\
+  {\cal D}_{2}^{\mu}d(x,y,z) u_{a}(x) 
\bar{\Psi}(z) G^{(5)}_{a\epsilon} \otimes \gamma^{\nu} \gamma_{\epsilon} \Psi(y)
%\nonumber \\
- {\cal D}_{1}^{\mu}d(x,y,z) u_{a}(y) 
\bar{\Psi}(z) G^{(5)}_{a\epsilon} \otimes \gamma^{\mu} \gamma_{\epsilon} \Psi(x) 
\nonumber \\
+  {\cal D}_{3}^{\mu}d(x,y,z) u_{a}(x) 
\bar{\Psi}(z) G^{(6)}_{a\epsilon} \otimes \gamma^{\mu} \gamma_{\epsilon} \Psi(y)
%\nonumber \\
- {\cal D}_{3}^{\mu}d(x,y,z) u_{a}(y) 
\bar{\Psi}(z) G^{(6)}_{a\epsilon} \otimes \gamma^{\nu} \gamma_{\epsilon} \Psi(x) 
\nonumber \\
+ {\cal D}_{1}^{\mu}d(x,y,z) u_{a}(z) 
\bar{\Psi}(x) G^{(7)}_{a\epsilon} \otimes \gamma^{\nu} \gamma_{\epsilon} \Psi(y)
%\nonumber \\
- {\cal D}_{2}^{\mu}d(x,y,z) u_{a}(z) 
\bar{\Psi}(y) G^{(7)}_{a\epsilon} \otimes \gamma^{\mu} \gamma_{\epsilon} \Psi(x) 
\nonumber \\
+ {\cal D}_{2}^{\mu}d(x,y,z) u_{a}(z) 
\bar{\Psi}(x) G^{(8)}_{a\epsilon} \otimes \gamma^{\nu} \gamma_{\epsilon} \Psi(y)
%\nonumber \\
- {\cal D}_{1}^{\mu}d(x,y,z) u_{a}(z) 
\bar{\Psi}(y) G^{(8)}_{a\epsilon} \otimes \gamma^{\nu} \gamma_{\epsilon} \Psi(x) 
\nonumber \\
+ {\cal D}_{3}^{\mu}d(x,y,z) u_{a}(z) 
\bar{\Psi}(x) G^{(9)}_{a\epsilon} \otimes \gamma^{\nu} \gamma_{\epsilon} \Psi(y)
%\nonumber \\
- {\cal D}_{3}^{\mu}d(x,y,z) u_{a}(z) 
\bar{\Psi}(y) G^{(9)}_{a\epsilon} \otimes \gamma^{\mu} \gamma_{\epsilon} \Psi(x)
\nonumber \\
+ d(x,y,z) 
[ u_{a}(x) \partial^{\mu}\bar{\Psi}(y) G^{(10)}_{a\epsilon} \otimes \gamma^{\nu}
\gamma_{\epsilon} \Psi(z)
%\nonumber \\
+ u_{a}(y) \partial^{\mu}\bar{\Psi}(x) G^{(10)}_{a\epsilon} \otimes \gamma^{\nu}
\gamma_{\epsilon} \Psi(z) ]
\nonumber \\
+ d(x,y,z) 
[ u_{a}(x) \bar{\Psi}(y) G^{(11)}_{a\epsilon} \otimes \gamma^{\nu}
\gamma_{\epsilon} \partial^{\mu}\Psi(z)
%\nonumber \\
+ u_{a}(y) \bar{\Psi}(x) G^{(11)}_{a\epsilon} \otimes \gamma^{\nu}
\gamma_{\epsilon} \partial^{\mu}\Psi(z) ]
\nonumber \\
+ d(x,y,z) 
[ u_{a}(x) \partial^{\mu}\bar{\Psi}(z) G^{(12)}_{a\epsilon} \otimes \gamma^{\nu}
\gamma_{\epsilon} \Psi(y)
%\nonumber \\
+ u_{a}(y) \partial^{\mu}\bar{\Psi}(z) G^{(12)}_{a\epsilon} \otimes \gamma^{\nu}
\gamma_{\epsilon} \Psi(x) ]
\nonumber \\
+ d(x,y,z) 
[ u_{a}(x) \bar{\Psi}(z) G^{(13)}_{a\epsilon} \otimes \gamma^{\nu}
\gamma_{\epsilon} \partial^{\mu}\Psi(y)
%\nonumber \\
+ u_{a}(y) \bar{\Psi}(z) G^{(13)}_{a\epsilon} \otimes \gamma^{\nu}
\gamma_{\epsilon} \partial^{\mu}\Psi(x) ]
\nonumber \\
+ d(x,y,z) u_{a}(z)
[ \partial^{\mu}\bar{\Psi}(x) G^{(14)}_{a\epsilon} \otimes \gamma^{\nu}
\gamma_{\epsilon} \Psi(y)
%\nonumber \\
+ \partial^{\mu}\bar{\Psi}(y) G^{(14)}_{a\epsilon} \otimes \gamma^{\nu}
\gamma_{\epsilon} \Psi(x) ]
\nonumber \\
+ d(x,y,z) u_{a}(z) 
[ \bar{\Psi}(x) G^{(15)}_{a\epsilon} \otimes \gamma^{\nu}
\gamma_{\epsilon} \partial^{\mu}\Psi(y)
%\nonumber \\
+ \bar{\Psi}(y) G^{(15)}_{a\epsilon} \otimes \gamma^{\nu}
\gamma_{\epsilon} \partial^{\mu}\Psi(x) ]
\nonumber\\
+ {\cal D}_{1}^{\rho}d(x,y,z)
u_{a}(x) \bar{\Psi}(y) G^{(16)}_{a\epsilon} \otimes 
\gamma^{\mu} \gamma^{\nu} \gamma_{\rho} \gamma_{\epsilon} \Psi(z)
\nonumber \\
- {\cal D}_{2}^{\rho}d(x,y,z)
u_{a}(y) \bar{\Psi}(x) G^{(16)}_{a\epsilon} \otimes 
\gamma^{\nu} \gamma^{\mu} \gamma_{\rho} \gamma_{\epsilon} \Psi(z)
\nonumber \\
+ {\cal D}_{2}^{\rho}d(x,y,z)
u_{a}(x) \bar{\Psi}(y) G^{(17)}_{a\epsilon} \otimes 
\gamma^{\mu} \gamma^{\nu} \gamma_{\rho} \gamma_{\epsilon} \Psi(z)
\nonumber \\
- {\cal D}_{1}^{\rho}d(x,y,z)
u_{a}(y) \bar{\Psi}(x) G^{(17)}_{a\epsilon} \otimes 
\gamma^{\nu} \gamma^{\mu} \gamma_{\rho} \gamma_{\epsilon} \Psi(z)
\nonumber \\
+ {\cal D}_{3}^{\rho}d(x,y,z)
u_{a}(x) \bar{\Psi}(y) G^{(18)}_{a\epsilon} \otimes 
\gamma^{\mu} \gamma^{\nu} \gamma_{\rho} \gamma_{\epsilon} \Psi(z)
\nonumber \\
- {\cal D}_{3}^{\rho}d(x,y,z)
u_{a}(y) \bar{\Psi}(x) G^{(18)}_{a\epsilon} \otimes 
\gamma^{\nu} \gamma^{\mu} \gamma_{\rho} \gamma_{\epsilon} \Psi(z) ]
\nonumber \\
+ {\cal D}_{1}^{\rho}d(x,y,z)
u_{a}(x) \bar{\Psi}(z) G^{(19)}_{a\epsilon} \otimes 
\gamma^{\mu} \gamma^{\nu} \gamma_{\rho} \gamma_{\epsilon} \Psi(y)
\nonumber \\
- {\cal D}_{2}^{\rho}d(x,y,z)
u_{a}(y) \bar{\Psi}(z) G^{(19)}_{a\epsilon} \otimes 
\gamma^{\nu} \gamma^{\mu} \gamma_{\rho} \gamma_{\epsilon} \Psi(x)
\nonumber \\
+ {\cal D}_{2}^{\rho}d(x,y,z)
u_{a}(x) \bar{\Psi}(z) G^{(20)}_{a\epsilon} \otimes 
\gamma^{\mu} \gamma^{\nu} \gamma_{\rho} \gamma_{\epsilon} \Psi(y)
\nonumber \\
- {\cal D}_{1}^{\rho}d(x,y,z)
u_{a}(y) \bar{\Psi}(z) G^{(20)}_{a\epsilon} \otimes 
\gamma^{\nu} \gamma^{\mu} \gamma_{\rho} \gamma_{\epsilon} \Psi(x)
\nonumber \\
+ {\cal D}_{3}^{\rho}d(x,y,z)
[ u_{a}(x) \bar{\Psi}(z) G^{(21)}_{a\epsilon} \otimes 
\gamma^{\mu} \gamma^{\nu} \gamma_{\rho} \gamma_{\epsilon} \Psi(y)
\nonumber \\
- u_{a}(y) \bar{\Psi}(z) G^{(21)}_{a\epsilon} \otimes 
\gamma^{\nu} \gamma^{\mu} \gamma_{\rho} \gamma_{\epsilon} \Psi(x) ]
\nonumber \\
+ {\cal D}_{1}^{\rho}d(x,y,z)
u_{a}(z) \bar{\Psi}(x) G^{(22)}_{a\epsilon} \otimes 
\gamma^{\mu} \gamma^{\nu} \gamma_{\rho} \gamma_{\epsilon} \Psi(y)
\nonumber \\
- {\cal D}_{2}^{\rho}d(x,y,z)
u_{a}(z) \bar{\Psi}(y) G^{(22)}_{a\epsilon} \otimes 
\gamma^{\nu} \gamma^{\mu} \gamma_{\rho} \gamma_{\epsilon} \Psi(x)
\nonumber \\
+ {\cal D}_{2}^{\rho}d(x,y,z)
u_{a}(z) \bar{\Psi}(x) G^{(23)}_{a\epsilon} \otimes 
\gamma^{\mu} \gamma^{\nu} \gamma_{\rho} \gamma_{\epsilon} \Psi(y)
\nonumber \\
- {\cal D}_{1}^{\rho}d(x,y,z)
u_{a}(z) \bar{\Psi}(y) G^{(23)}_{a\epsilon} \otimes 
\gamma^{\nu} \gamma^{\mu} \gamma_{\rho} \gamma_{\epsilon} \Psi(x)
\nonumber \\
+ {\cal D}_{3}^{\rho}d(x,y,z)
u_{a}(z) \bar{\Psi}(x) G^{(24)}_{a\epsilon} \otimes 
\gamma^{\mu} \gamma^{\nu} \gamma_{\rho} \gamma_{\epsilon} \Psi(y)
\nonumber \\
- {\cal D}_{3}^{\rho}d(x,y,z) u_{a}(z)
\bar{\Psi}(y) G^{(24)}_{a\epsilon} \otimes 
\gamma^{\nu} \gamma^{\mu} \gamma_{\rho} \gamma_{\epsilon} \Psi(x) \}
\nonumber\\
- \{\mu \longleftrightarrow \nu\}
\eea
\newpage
We get from the equation (\ref{co3})
\bea
D_{5}^{\emptyset\emptyset[\mu]}(x,y,z)_{0} = 
(\bar{s}B_{5})^{\emptyset\emptyset[\mu]}(x,y,z)
+ D_{5\delta}^{\emptyset\emptyset[\mu]}(x,y,z)
\nonumber
\eea
we obtain the following system
\bea
- F_{40} + F_{46} + F_{48} - G_{21} = 0
\nonumber \\
F_{40} + F_{47} - G_{19} - G_{20} = {1\over 2} t_{1}
\nonumber \\
- F_{41} - F_{42} + F_{47} + G_{21} = 0
\nonumber \\
F_{40} + F_{42} + F_{44} + G_{24} = - {1\over 2} t_{1}
\nonumber \\
F_{41} + F_{43} + F_{45} - G_{24} = {1\over 2} t_{1}
\nonumber \\
- F_{41} + F_{44} - G_{22} - G_{23} =  2 t_{2} - t_{3}
\nonumber \\
- F_{43} - F_{47} - F_{48} - G_{18} = 0
\nonumber \\
F_{43} - F_{46} - G_{16} - G_{17} = - {1\over 2} t_{1}
\nonumber \\
- F_{44} - F_{45} - F_{46} + G_{18} = 0
\nonumber \\
- F_{1} + F_{20} - F_{24} + 2 F_{40} - 2 F_{48} + G_{6} = 0
\nonumber \\
F_{1} + F_{22} - 2 F_{40} - G_{4} + G_{5} + 2 G_{19} = 0
\nonumber \\
- F_{2} - F_{23} + F_{25} + G_{21} = 0
\nonumber \\
F_{2} + F_{26} + G_{4} + G_{20} = 0
\nonumber \\
F_{3} - F_{5} - F_{22} + 2 F_{42} - 2 F_{47} - G_{6} = 0
\nonumber \\
- F_{4} - F_{8} - F_{19} - F_{25} - 2 F_{41} = 0
\nonumber \\
F_{4} - F_{9} - F_{26} - 2 F_{42}  + G_{6} = 0
\nonumber \\
- F_{6} + F_{8} - F_{21} + 2 F_{41} - G_{21} = 0
\nonumber \\
F_{6} + F_{9} + 2 F_{42} + G_{5} - G_{20} = 0
\nonumber \\
- F_{7} + F_{19} - F_{27} - 2 F_{40} - G_{6} = 0
\nonumber \\
F_{7} + F_{21} + 2 F_{40} - G_{5} - G_{19} = 0
\nonumber \\
F_{23} + F_{27} - G_{4} + G_{19} = 0
\nonumber \\
- F_{1} - F_{7} - F_{10} - F_{16} = 0
\nonumber \\
F_{1} - F_{9} - F_{17} + G_{9} = 0
\nonumber \\
F_{2} - F_{6} - F_{13} - 2 F_{42} - 2 F_{44} - G_{9} = t_{1}
\nonumber \\
- F_{3} - F_{14} + F_{16} + G_{24} = t_{4}
\nonumber \\
F_{3} + F_{17} + G_{7} + G_{23} = - 4 t_{2}
\nonumber \\
- F_{4} + F_{11} - F_{15} - 2 F_{41} - 2 F_{45} + G_{9} = 0
\nonumber \\
F_{4} + F_{13} + 2 F_{41} - G_{7} + G_{8} + 2 G_{22} = 2 t_{3}
\nonumber \\
- F_{5} + F_{7} - F_{12} - G_{24} = t_{1} + t_{4}
\nonumber \\
F_{5} + F_{9} + G_{8} - G_{23} = 0
\nonumber \\
- F_{8} + F_{10} - F_{18} - G_{9} = t_{1}
\nonumber \\
F_{8} + F_{12} - G_{8} - G_{22} = - 2 t_{3}
\nonumber \\
F_{14} + F_{18} - G_{7} + G_{22} = 0
\nonumber \\
- F_{10} + F_{21} - F_{23} + 2 F_{43} + 2 F_{48} + G_{3} = 0
\nonumber \\
F_{10} + F_{19} - 2 F_{43} - G_{1} + G_{2} + 2 G_{16} = t_{1}
\nonumber \\
- F_{11} - F_{24} + F_{26} + 2 F_{47} + G_{18} = 0
\nonumber \\
F_{11} + F_{25} + 2 F_{46} + G_{1} + G_{17} = t_{1}
\nonumber \\
F_{12} - F_{14} - F_{19} + 2 F_{45} + 2 F_{46} - G_{3} = 0
\nonumber \\
- F_{13} - F_{17} - F_{22} - F_{26} - 2 F_{44} - 2 F_{47} = 0
\nonumber \\
F_{13} - F_{18} - F_{25} - 2 F_{45} - 2 F_{46} + G_{3} = 0
\nonumber \\
- F_{15} + F_{17} - F_{20} + 2 F_{44} - G_{18} = 0
\nonumber \\
F_{15} + F_{18} + 2 F_{45} + G_{2} - G_{17} = 0
\nonumber \\
- F_{16} + F_{22} - F_{27} - 2 F_{43} - 2 F_{48} - G_{3} = 0
\nonumber \\
F_{16} + F_{20} + 2 F_{43} - G_{2} - G_{16} = - t_{1}
\nonumber \\
F_{24} + F_{27} + 2 F_{48} - G_{1} + G_{16} = 0
\nonumber \\
- F_{2} - F_{29} + F_{33} = 0
\nonumber \\
- F_{4} - F_{33} + G_{14} = 0
\nonumber \\
- F_{6} + F_{29} - G_{14} = 0
\nonumber \\
- F_{7} - F_{28} + F_{32} - 2 F_{40} = 0
\nonumber \\
- F_{8} - F_{32} - 2 F_{41} - G_{14} = 0
\nonumber \\
- F_{9} + F_{28} - 2 F_{42} + G_{14} = 0
\nonumber \\
F_{11} - F_{31} + F_{35} + 2 F_{43} = 0
\nonumber \\
F_{13} + F_{31} + 2 F_{44} - G_{15} = 0
\nonumber \\
F_{15} - F_{35} + 2 F_{45} + G_{15} = 0
\nonumber \\
F_{16} - F_{30} + F_{34} = 0
\nonumber \\
F_{17} + F_{30} + G_{15} = 0
\nonumber \\
F_{18} - F_{34} - G_{15} = 0
\nonumber \\
F_{19} - F_{32} - G_{10} = 0
\nonumber \\
F_{21} + F_{32} - F_{36} = 0
\nonumber \\
F_{23} + F_{36} + G_{10} = 0
\nonumber \\
F_{25} - F_{33} + G_{10} = 0
\nonumber \\
F_{26} + F_{33} - F_{37} = 0
\nonumber \\
F_{27} + F_{37} - G_{10} = 0
\nonumber \\
F_{34} - F_{38} + G_{3} + 2 G_{18} = 0
\nonumber \\
- F_{34} - G_{2} - G_{11} + 2 G_{17} = 0
\nonumber \\
F_{35} - F_{39} - G_{18} = 0
\nonumber \\
- F_{35} + G_{11} - G_{17} = 0
\nonumber \\
F_{38} - G_{1} + G_{11} + 2 G_{16} = 0
\nonumber \\
F_{39} - G_{11} - G_{16} = 0
\nonumber \\
F_{20} + F_{30} + F_{39} + 2 F_{46} = 0
\nonumber \\
F_{22} - F_{30} + 2 F_{47} - G_{13} = 0
\nonumber \\
F_{24} - F_{39} + 2 F_{48} + G_{13} = 0
\nonumber \\
F_{25} + F_{31} + F_{38} = 0
\nonumber \\
F_{26} - F_{31} + G_{13} = 0
\nonumber \\
F_{27} - F_{38} - G_{13} = 0
\nonumber \\
F_{28} + F_{37} - G_{6} = 0
\nonumber \\
- F_{28} - G_{5} - G_{12} = 0
\nonumber \\
F_{29} + F_{36} + G_{21} = 0
\nonumber \\
- F_{29} + G_{12} + G_{20} = 0
\nonumber \\
- F_{36} - G_{12} + G_{19} = 0
\nonumber \\
- F_{37} - G_{4} + G_{12} = 0
\nonumber \\
- F_{31} - F_{33} = 0
\nonumber \\
- F_{37} - G_{13} = 0
\nonumber \\
F_{38} - G_{10} = 0
\label{F}
\eea

We can solve this system {\it iff} we impose
\be
t_{1} - 2 t_{2} - t_{3} + t_{4} = 0.
\ee
\newpage
\subsection{The Final Result}
Collecting the results from the preceding Subsections we obtain the following result:
\begin{thm}
The equation
\be
D^{IJK}_{\rm triangle}(x,y,z)_{0} = (\bar{s}B)^{IJK}(x,y,z) 
+ D^{IJK}_{\rm triangle}(x,y,z)_{\delta}
\ee
is true iff the following restrictions are true:
\bea
f^{(3)}_{abc} = 6 f^{(0)}_{abc}
\nonumber\\
f^{(4)}_{abc} = 8 f^{(0)}_{abc}
\nonumber\\
t_{1} - 2 t_{2} - t_{3} + t_{4} = 0.
\eea
\end{thm}
To complete the proof we have to prove first a similar result for the 1PR
contributions, 
namely:
\be
D^{IJK}_{\rm 1PR}(x,y,z)_{0} = (\bar{s}b)^{IJK}(x,y,z) 
+ D^{IJK}_{\rm 1PR}(x,y,z)_{\delta}
\ee
is true without other restrictions.

It can be seen that both delta-contributions (obtained from the triangle and 
1PR contributions) are non-trivial, i.e. they cannot be made null. If we apply 
the relations of the type (\ref{on-shell}) we will get two type of terms:
(a) terms with derivatives on the delta distribution and (b) terms without derivatives
on the delta distribution. We can write  contribution (a) as a coboundary plus a 
contribution (b). So in the end we have to check that the remaining contribution
(b) is null.
This follows by direct computations. This means that we have 
\begin{thm}
The equation
\be
D^{IJK}(x,y,z)_{(1)} = (\bar{s}B)^{IJK}(x,y,z) + {\rm
super-renormalizable~terms}
\ee
is true iff the following restrictions are true:
\bea
f^{(3)}_{abc} = 6 f^{(0)}_{abc}
\nonumber\\
f^{(4)}_{abc} = 8 f^{(0)}_{abc}
\nonumber\\
t_{1} - 2 t_{2} - t_{3} + t_{4} = 0.
\label{restrictions}
\eea
\end{thm}

The two-loop contribution can be analysed in the same way and does not bring new 
constraints. This is our final result.
\section{Conclusions}
We have proved that the super-renormalizability property is true for Yang-Mills models 
in the third order of the perturbation theory if we have the three relations from above.
We have checked that the electro-weak sector does not fulfill them so we must look
for another gauge group having two properties: it should lead to a solution of the
preceding equations and it should be in agreement with the phenomenology (not very
``far" from the standard model). This problem will be addressed in further
publications.
However, let us mention that the second relation (\ref{restrictions}) gives in
the QCD sector that the number of colors must be $3$.
We must also investigate if the super-renormalizability property can be implemented 
in arbitrary orders of the perturbation theory and try to extend the result for 
gravity also.


\newpage
\begin{thebibliography}{99}

\bibitem{BS}
N. N. Bogoliubov, D. Shirkov,
``{\it Introduction to the Theory of Quantized Fields}",
John Wiley and Sons, 1976 (3rd edition)

\bibitem{DF}
M. D\"utsch, K. Fredenhagen,
``{\it A Local (Perturbative) Construction of Observables in Gauge Theories:
the Example of QED}", Commun. Math. Phys. {\bf 203} (1999) 71-105

\bibitem{EG}
H. Epstein, V. Glaser,
``{\it The R\^ole of Locality in Perturbation Theory}",
Ann. Inst. H. Poincar\'e {\bf 19 A} (1973) 211-295

\bibitem{Gl}
V. Glaser,
``{\it Electrodynamique Quantique}",
L'enseignement du 3e cycle de la physique en Suisse Romande (CICP), Semestre
d'hiver 1972/73

\bibitem{YM} D. R. Grigore
``{\it On the Uniqueness of the Non-Abelian Gauge Theories in Epstein-Glaser 
Approach to Renormalisation Theory}", 
Romanian J. Phys. {\bf 44} (1999) 853-913

\bibitem{standard} D. R. Grigore
``{\it The Standard Model and its Generalisations in Epstein-Glaser 
Approach to Renormalisation Theory}", 
Journ. Phys. {\bf A 33} (2000) 8443-8476 

\bibitem{fermi} D. R. Grigore
``{\it The Standard Model and its Generalisations in Epstein-Glaser 
Approach to Renormalisation Theory II: the Fermion Sector and the Axial
Anomaly}", \\
Journ. Phys {\bf A 34} (2001) 5429-5462

\bibitem{cohomology}
D. R. Grigore, 
``{\it Cohomological Aspects of Gauge Invariance in the Causal Approach}",
Romanian Journ. Phys. {\bf 55} (2010) 386-438

\bibitem{super2} D. R. Grigore
``{\it On the Super-Renormalizablity of Gauge Models in the Causal Approach}",
hep-th/1301.2893

\bibitem{PS}
G. Popineau, R. Stora, 
``{\it A Pedagogical Remark on the Main Theorem of Perturbative Renormalization
Theory}", unpublished preprint

\bibitem{Sc1}
G. Scharf,
``{\it Finite Quantum Electrodynamics: The Causal Approach}",
(second edition) Springer, 1995

\bibitem{Sc2}
G. Scharf,
``{\it Quantum Gauge Theories. A True Ghost Story}",
John Wiley, 2001
and ``{\it Quantum Gauge Theories - Spin One and Two}",
Google books, 2010

\bibitem{Sto1}
R. Stora,
``{\it Lagrangian Field Theory}",
Les Houches lectures, Gordon and Breach, N.Y., 1971, 
C. De Witt, C. Itzykson eds.

\bibitem{St1}
O. Steinmann,
``{\it Perturbation Expansions in Axiomatic Field Theory}",
Lect. Notes in Phys. {\bf 11}, Springer, 1971

\end{thebibliography}
\end{document}